\RequirePackage{fix-cm}
\RequirePackage{fixltx2e}
\documentclass[11pt,aip, preprint]{revtex4-1}
\usepackage{lmodern}
\usepackage[T1]{fontenc}
\usepackage[utf8]{inputenc}
\usepackage{geometry}
\usepackage{color}
\usepackage{float}
\usepackage{amsmath}
\usepackage{amssymb}
\usepackage{graphicx}
\PassOptionsToPackage{version=3}{mhchem}
\usepackage{mhchem}
\usepackage{wasysym}
\usepackage{microtype}
\usepackage[unicode=true,
 bookmarks=false,
 breaklinks=true,pdfborder={0 0 0},pdfborderstyle={},backref=false,colorlinks=true]
 {hyperref}
\hypersetup{pdftitle={Comparison of methods},
 citecolor =blue, linkcolor = blue,  urlcolor  = blue}

\makeatletter

\providecommand{\tabularnewline}{\\}
\newcommand{\lyxdot}{.}

\floatstyle{ruled}
\newfloat{algorithm}{H}{loa}
\providecommand{\algorithmname}{Algorithm}
\floatname{algorithm}{\protect\algorithmname}

\usepackage{datetime}
\usepackage{refstyle}
\usepackage{longtable}
\usepackage{url}
\usepackage[title,toc,page,header]{appendix} 

\makeatother

\begin{document}

\title{Probabilistic performance estimators for computational chemistry
methods: Systematic Improvement Probability and Ranking Probability
Matrix. I. Theory}

\author{Pascal PERNOT}

\affiliation{Institut de Chimie Physique, UMR8000, CNRS, Université Paris-Saclay,
91405 Orsay, France}
\email{Pascal.Pernot@universite-paris-saclay.fr
}

\author{Andreas SAVIN }

\affiliation{Laboratoire de Chimie Théorique, CNRS and UPMC Université Paris 06,
Sorbonne Universités, 75252 Paris, France}
\email{Andreas.Savin@lct.jussieu.fr}

\begin{abstract}
The comparison of benchmark error sets is an essential tool for the
evaluation of theories in computational chemistry. The standard ranking
of methods by their Mean Unsigned Error is unsatisfactory for several
reasons linked to the non-normality of the error distributions and
the presence of underlying trends. Complementary statistics have recently
been proposed to palliate such deficiencies, such as quantiles of
the absolute errors distribution or the mean prediction uncertainty.
We introduce here a new score, the systematic improvement probability
(SIP), based on the direct system-wise comparison of absolute errors.
Independently of the chosen scoring rule, the uncertainty of the statistics
due to the incompleteness of the benchmark data sets is also generally
overlooked. However, this uncertainty is essential to appreciate the
robustness of rankings. In the present article, we develop two indicators
based on robust statistics to address this problem: $P_{inv}$, the
inversion probability between two values of a statistic, and $\mathbf{P}_{r}$,
the ranking probability matrix. We demonstrate also the essential
contribution of the correlations between error sets in these scores
comparisons.
\end{abstract}
\maketitle

\section{Introduction}

Benchmarks are a central tool for the evaluation of new theories/methods
in quantum chemistry \citep{Mata2017}. Amongst many possible metrics
\citep{Civalleri2012}, the most common benchmarking statistics are
the mean unsigned error (MUE/MAD/MAE...), mean signed error (MSE),
root mean squared error (RMSE) and root mean squared deviation (RMSD).
The explicit definition of these scores is given in a previous article
\citep{Pernot2018}. In a vast majority of benchmark studies, the
MUE, or some variant of it, is used to compare methods performance.
Recently \citep{Pernot2018}, we proposed a more informative probabilistic
score, the 95th percentile of the absolute errors distribution ($Q_{95}$).\footnote{We argued that $Q_{95}$ is more informative than the MUE, because
the latter provides probabilistic information only if the errors distribution
is zero-centered normal, a rather unlikely occurrence. In contrast,
$Q_{95}$ gives us the error level that one has only 5\,\% chance
to exceed in a new calculation (provided that the reference dataset
is representative of the systems for which predictions are sought).
The end-users can easily check if this threshold meets their expectations.
We recently realized that the 90th percentile (noted $P_{90}$) has
been used by Thakkar and colleagues in the same spirit \citep{Thakkar2015,Wu2015b}.
We think $Q_{95}$ is more appropriate because of its direct link
to the enlarged uncertainty $u_{95}$ recommended in the thermochemistry
literature \citep{Ruscic2014,Pernot2018}.} 

Whichever the statistic used, the question remains of the robustness
of such scores and rankings with respect to the choice of the reference
dataset. One easily conceives that the values of these statistics
change unpredictably when one adds or removes points in the dataset.
Benchmarks implicitly assume that the error sets are representative
samples of unknown distributions characterizing model errors for each
method -- the more systems in the dataset, the best the approximation
of the underlying distributions. The quest for large datasets incurs
heavy computer charges to perform benchmarks, and there is also a
trend to reduce this burden by looking for small, optimally representative,
datasets \citep{Gould2018,Morgante2019}. Besides, there are several
properties for which the reference data are rather sparse, leading
to rather small datasets. Another trend, enhanced by the development
of machine learning is to replace experimental values by gold standard
calculations, with limitations on the size of accessible systems \citep{Ramakrishnan2015,Zaspel2019}.
As the estimated values of the statistics and their uncertainties
depend on the size of the dataset, it is important to assess this
size effect and its impact on statistics comparison and ranking. 

This question has been considered recently by Proppe and Reiher \citep{Proppe2017},
who used bootstrapping to assess the impact of dataset size and reference
data uncertainty on the first place in an intercomparison of M\"ossbauer
isomer shifts estimated by a dozen of DFAs. They concluded that for
their dataset of $N=39$ values, at least three methods were competing
for the first place, with a slight probabilistic advantage for PBE0.
This is a very interesting contribution to the quality assessment
of benchmarking tools. We recently considered another approach to
this problem by defining an inversion probability $P_{inv}$ for the
ranking of two methods \citep{Pernot2018}. Our definition, which
was based on the assumption of a normal distribution of statistics
differences and neglected error sets correlations, deserves a more
general setup. 

In the present study, we revisit the ranking uncertainty problem along
several complementary lines:
\begin{enumerate}
\item we consider the statistical significance of the difference between
two values of a statistic: it depends both on the uncertainty on the
estimated values, which is notably influenced by the dataset size,
and on the correlation between these values, which is due in a large
part to the use of a common reference dataset \citep{Nicholls2016}.
A few specific points have also to be considered: the non-normality
of the error sets distributions, the small size of some datasets,
the uncertainty on reference data, and some properties of quantiles
estimators. 
\item we define a ranking probability matrix $\mathrm{P}_{r}$, generalizing
the proposition of Proppe and Reiher \citep{Proppe2017}, which enables
us to propose an efficient visual assessment of the robustness of
rankings.
\item we introduce a new statistic (the systematic improvement probability,
SIP) that conveys the proportion of systems in the benchmark data
set for which one method has smaller absolute errors than the other,
and the expected gain or loss when switching between methods.
\end{enumerate}
\medskip{}

The article is structured as follows. In Section\,\ref{subsec:Error-sets,-their},
we consider the uncertainty and correlations of the error sets used
in benchmarking, and in Section\,\ref{subsec:Statistics,-their-uncertainty}
how these are transferred to benchmarking statistics. Correlation
of error sets and their statistics are central to the developments
presented next: Section\,\ref{subsec:Error-sets,-their} introduces
the SIP, based on the system-wise comparison of absolute errors, and
Section\,\ref{subsec:Pair-wise-comparison-of-1} develops bootstrap-based
tools to compare uncertain and correlated statistics, leading to the
ranking inversion probability $P_{inv}$ and ranking probability matrix
$\mathrm{P}_{r}$. Implementation details are reported in Section\,\ref{subsec:Implementation}.
Section\,\ref{sec:Conclusions} provides a brief conclusion, but
a detailed discussion is deferred to Paper\,II \citep{Pernot2020a},
where these methods are applied to nine datasets taken from the recent
benchmarking literature and covering a wide range of dataset sizes
and properties. 

\section{Error sets, their uncertainty and correlation\label{subsec:Error-sets,-their}}

Benchmarking of a method $M$ is based on the statistical analysis
of its error set ($E_{M}=\left\{ e_{i}(M)\right\} _{i=1}^{N}$), based
on a set of $N$ calculated ($C_{M}=\left\{ c_{i}(M)\right\} _{i=1}^{N}$)
and reference data ($R=\left\{ r_{i}\right\} _{i=1}^{N}$), where
\begin{equation}
e_{i}(M)=r_{i}-c_{i}(M)\label{eq:errors-def}
\end{equation}

\subsubsection{Uncertainty}

As the reference data or even the calculated values can be uncertain,
one should consider that the error sets contain uncertain values when
estimating and comparing statistics. Experimental or computational
uncertainties being typically estimated by standard deviations, one
can use the method of combination of variances to get the uncertainty
on the errors \citep{GUM},
\begin{equation}
u(e_{i})=\sqrt{u(r_{i})^{2}+u(c_{i})^{2}}\label{eq:ue-def}
\end{equation}
where $u(x)$ is the uncertainty on $x$. This formula assumes that
the individual errors on the reference data and calculated values
are uncorrelated. For an experimental reference value $r_{i}$, $u(r_{i})$
would typically be a measurement uncertainty. For a computed reference
value $r_{i}$ and for a calculated value $c_{i}$, uncertainty might
come from numerical uncertainty due to the use of finite precision
arithmetics and discretization errors \citep{Janes2011,Cances2017},
statistical uncertainty (\emph{e.g.}, for Monte Carlo methods \citep{Reynolds_1982,Cailliez2011}),
or parametric uncertainty (\emph{e.g.}, for calibrated methods \citep{Mortensen2005,Cailliez2011,Pernot2017b,Bakowies2019,Bakowies2020}).

We consider here deterministic computational chemistry methods for
which the sole uncertainty source is arithmetic uncertainty, assumed
to be well controlled. The uncertainty on errors is then equal to
the reference data uncertainty $u(e_{i})\equiv u(r_{i})$. For the
sake of generality, the $u(e_{i})$ notation is preserved in the following.

\subsubsection{Error sets covariance and correlation}

Let us consider a set of $K$ methods $\left\{ M_{i}\right\} _{i=1}^{K}$.
The covariance \citep{Snedecor1989} of the error sets for two method
can be decomposed as
\begin{align}
\mathrm{cov}(E_{i},E_{j}) & =\mathrm{cov}(R-C_{i},R-C_{j})\\
 & =\mathrm{var}(R)+\mathrm{cov}(C_{i},C_{j})-\mathrm{cov}(R,C_{i})-\mathrm{cov}(R,C_{j})
\end{align}
where, for brevity, we use shortened notations such as $E_{i}\equiv E_{M_{i}}$.
It is not possible to predict the sign and amplitude of  $\mathrm{cov}(E_{i},E_{j})$
from this decomposition, but a few considerations on the various terms
might be helpful:
\begin{itemize}
\item when comparing computational chemistry methods, it is very likely
that their prediction sets are strongly positively correlated (covariant).
It is also very likely that the predictions of good methods have a
strong positive covariance with the reference data, if the latter
are not dominated by measurement errors. Besides, one can expect that
the variance of the reference data set is of the same order (possibly
larger if there are notable experimental errors) as the variance/covariances
of the calculated data set. So, in a typical comparison scenario,
$\mathrm{cov}(E_{i},E_{j})$ results from the compensation of terms
with similar magnitudes, and one should not expect a null covariance
of error sets.
\item if reference data uncertainties are larger than prediction errors,
the covariance should be dominated by $\mathrm{var}(R)$, and all
error sets should be strongly positively correlated.
\end{itemize}
Instead of covariances, it is easier to work with the correlation
coefficients between error sets (normalized covariances)
\begin{align}
\mathrm{cor}(E_{i},E_{j}) & =\frac{\mathrm{cov}(E_{i},E_{j})}{\sigma_{E_{i}}\sigma_{E_{j}}}
\end{align}
where $\sigma_{E_{i}}$is the standard deviation of the error set
$E_{i}$, assumed finite. We will show in Paper\,II \citep{Pernot2020a}
through case studies that the correlation matrix contains relevant
information on the quality of datasets and the proximity of methods.

\subsubsection{Representation}

\begin{figure}[!tb]
\noindent \begin{centering}
\includegraphics[viewport=200bp 250bp 1800bp 1800bp,clip,width=0.45\textwidth]{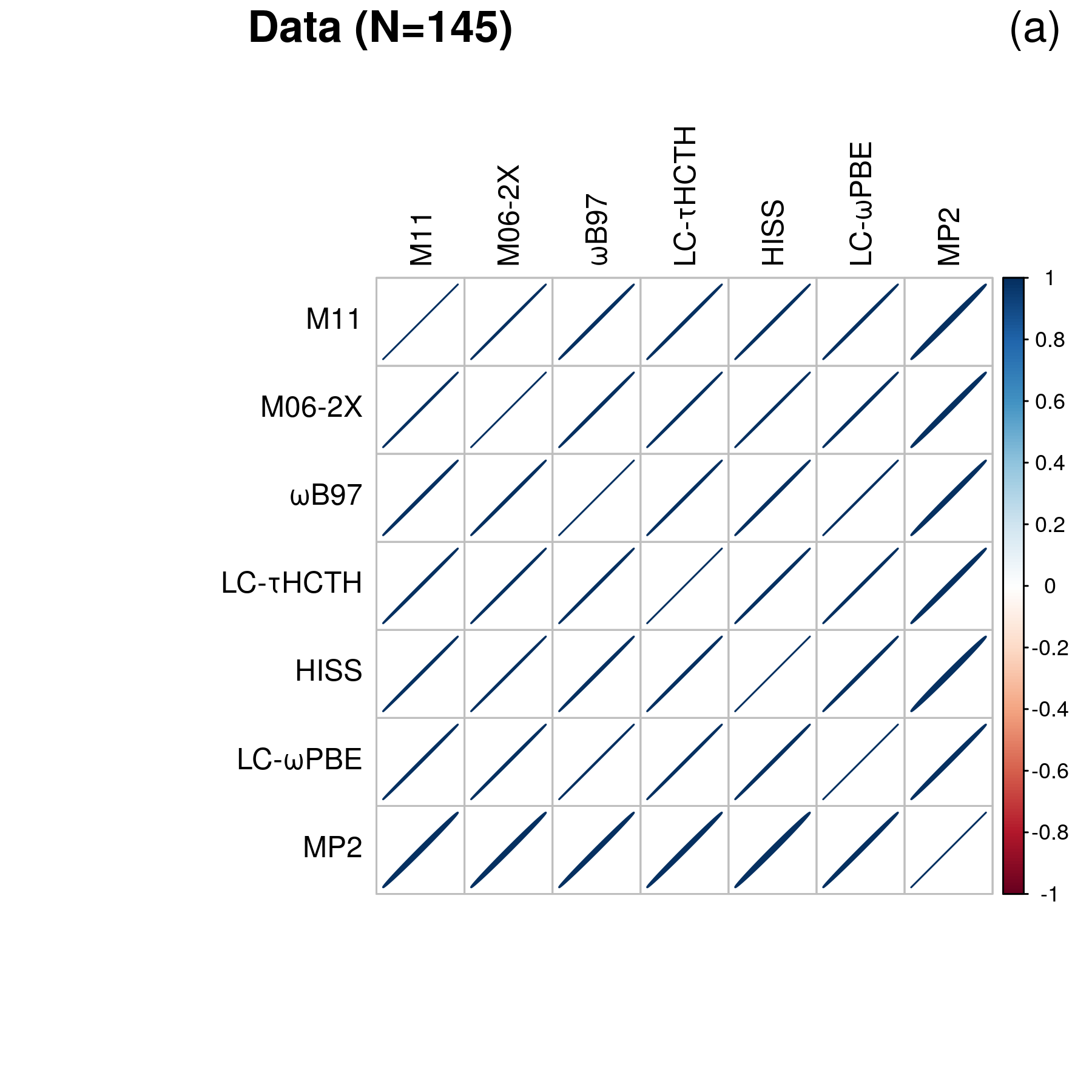}\includegraphics[viewport=200bp 250bp 1800bp 1800bp,clip,width=0.45\textwidth]{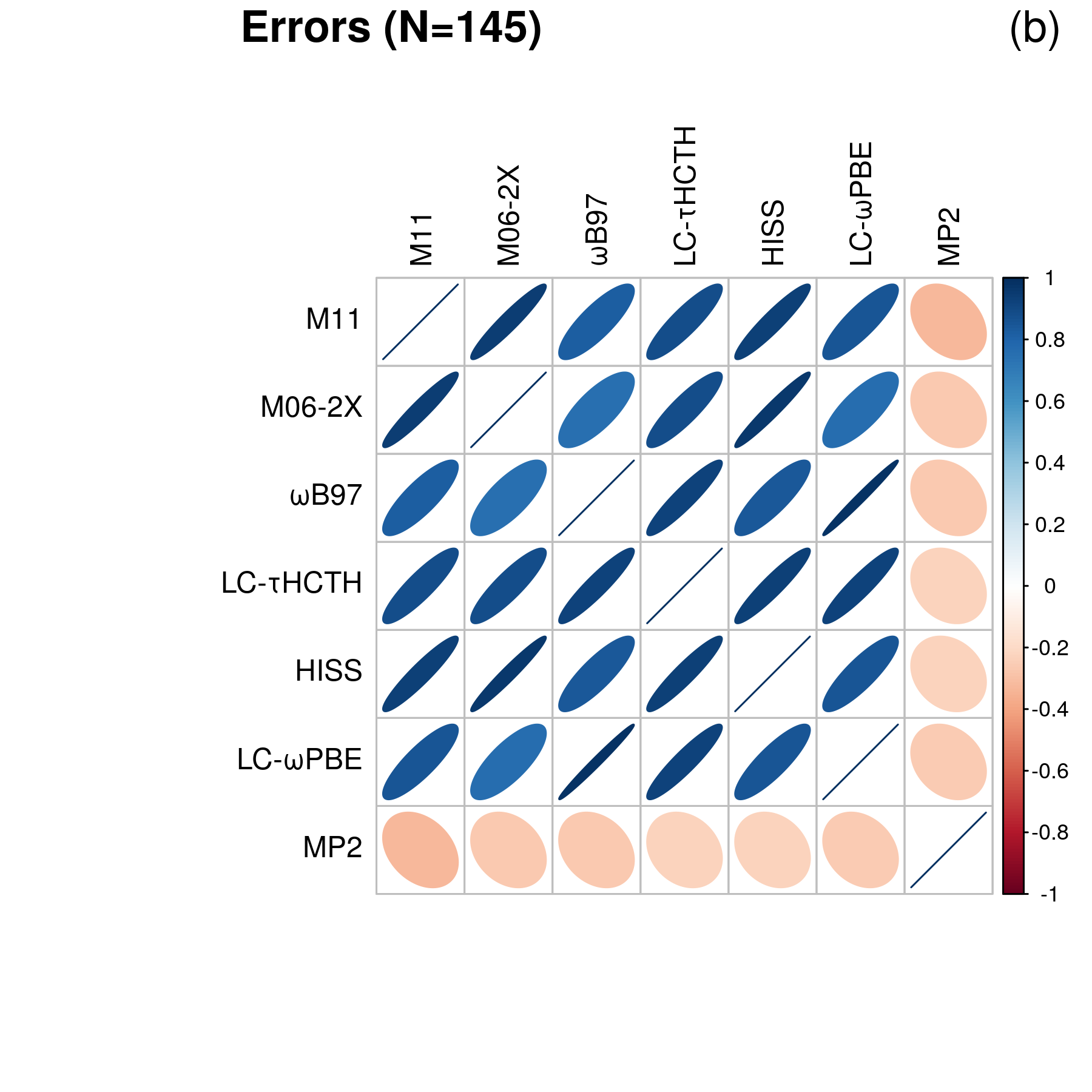}
\par\end{centering}
\noindent \centering{}\caption{\label{fig:cmat-example}Rank correlation matrices between (a) data
sets and (b) errors sets of polarizabilities for case WU2015 (Paper\,II
\citep{Pernot2020a}).}
 
\end{figure}

Correlation matrices can be represented by combining a color scheme
and an ellipse model \citep{Murdoch1996} (Fig.\,\ref{fig:cmat-example}),
such that a blue right-slanted ellipse stands for a positive correlation,
a red left-slanted ellipse for a negative one, and a white (invisible)
disk for a null correlation. The larger the absolute value of the
correlation, the darker the color and the thinner the ellipse. 

For the example showcased in Fig.\,\ref{fig:cmat-example}(a), one
sees that all the datasets $C_{i}$ are all strongly positively correlated,
meaning that all methods produce closely the same trend. By contrast,
the error sets $E_{i}$ present a more relaxed pattern (Fig.\,\ref{fig:cmat-example}(b)),
with weaker positive correlations, and even a very small negative
correlation for MP2 with all the other error sets. Having noticed
this, one can remark that MP2 data present also smaller correlation
coefficients with other datasets, although this is barely visible
on the figure (the difference bears on the third digit of the correlation
coefficients). In the following, we present correlation matrices for
error sets only.

\section{Statistics, their uncertainty and correlation\label{subsec:Statistics,-their-uncertainty}}

\subsubsection{Uncertainty}

The value $s$ of a statistic $S$ (MSE, MUE, $Q_{95}$...) estimated
on an error set is generally uncertain, with uncertainty estimated
by its standard error $u(s)$. Two main uncertainty sources should
be considered: (1) the limited size $N$ of the reference data sample,
and (2) the uncertainty on errors, $u(e_{i})$ (Section\,\ref{subsec:Error-sets,-their}).
Unless the dataset is exhaustive (\emph{e.g.}, a dataset containing
a property for a complete class of systems), the first source is always
present. For experimental reference data, the second source is also
always present, but experimental uncertainty is rarely available for
large datasets, and a common practice seems to be to ignore them in
the statistical analysis (although they are often discussed to assess
the quality of the dataset). Some studies considered the effect of
representative uncertainty levels on benchmarking conclusions \citep{Pernot2015,DeWaele2016,Proppe2017}.

In Appendix\,\ref{sec:Estimation-of-the}, the impact of both uncertainty
sources is illustrated on the mean value (MSE), for which analytical
formulae are available. The strategy to handle reference data uncertainty
depends on their distribution. If the reference data uncertainties
are uniform over the dataset, the hypothesis of \emph{i.i.d.} errors
holds, and standard statistical procedures can be applied (unless
one is interested in quantifying specifically model errors \citep{Pernot2015,Proppe2017}).
Otherwise, weighted statistics have to be used \citep{Pernot2015,Proppe2017},
which will not be considered here. Instead, we assume that datasets
should not include data with extreme uncertainty values.

Simple formulae for standard errors, such as those for the mean (a
linear statistic), are not available for non-linear statistics such
as the MUE or $Q_{95}$. Moreover, in order to avoid some of the limitations
implied by such formulae (\emph{e.g.}, normality hypothesis), one
can use a general method to estimate the standard error of any statistic:
the bootstrap \citep{Efron1979,Efron1991,Hesterberg2015}. It is a
Monte Carlo sampling method which consists in random draws with replacement
of $N'$ values from a dataset of size $N$. In the standard bootstrap,
one uses $N'=N$, \emph{i.e.}, the generated samples have the same
size as the original set. The bootstrap has been shown to provide
reliable estimation of uncertainty, but the mean values unavoidably
reflect the bias due to the original data set \citep{Hesterberg2015}.
In consequence, we estimate in the following the mean values from
the original sample and the uncertainties from the bootstrap samples.
The main limitation of the bootstrap is its hypothesis of $i.i.d.$
data, but it is consistent with our choice to avoid reference datasets
with a large uncertainty range. 

\subsubsection{Correlation}

The statistics covariance $\mathrm{cov}(s_{1},s_{2})$ derives from
the mathematical expression of $S$ and from the variances and covariance
of the error sets, $\mathrm{cov}(E_{1},E_{2})$. To estimate $\mathrm{cov}(s_{1},s_{2})$
in the case of a linear statistic, one can directly apply the generalization
of the combination of variances to several model outputs \citep{GUM-Supp2}.
For the MSE, it is easy to demonstrate that the covariance is transferred
in totality: $\mathrm{cov}(\overline{e}_{1},\overline{e}_{2})=\mathrm{cov}(E_{1},E_{2})$,
where $\bar{x}$ is the mean value of $X$. More generally, for linear
statistics, $\mathrm{cov}(E_{1},E_{2})=0\Longrightarrow\mathrm{cov}(s_{1},s_{2})=0$.
For non-linear statistics, such as the MUE or $Q_{95}$, the combination
of covariances is unsuitable, and Monte Carlo strategies are used. 

To illustrate the transfer of correlation from error sets to non-linear
statistics, we performed a Monte Carlo study, detailed in Appendix\,\ref{sec:Covariance-of-scores},
with scenarii implying diverse distribution shapes. A few trends can
be derived from this study, notably that for the MUE and $Q_{95},$
$\mathrm{cor}(s_{1},s_{2})$ is a convex, positive function of $\mathrm{cor}(E_{1},E_{2})$.
Moreover, for a given value of $\mathrm{cor}(E_{1},E_{2})$ one observes
that $\mathrm{cor}(MUE_{1},MUE_{2})\ge\mathrm{cor}(Q_{95,1},Q_{95,2})$.
As we explored only a fraction of the possible scenarii for the errors
distributions, these trends should not be considered as general. Our
main point is that the correlation of error sets is at least partially
transferred to the derived statistics, a fact to be considered when
comparing the values of these statistics.

\section{Pair-wise comparison of errors\label{subsec:Pair-wise-comparison-of}}

We define the systematic improvement probability (SIP) between two
methods $M_{i}$ and $M_{j}$ as the proportion of systems in the
reference set for which the absolute error decreases when using $M_{i}$
instead of $M_{j}$. It is estimated as
\begin{align}
\mathrm{SIP}_{i,j} & =\frac{D_{i,j}}{N}\\
D_{i,j} & =\sum_{k=1}^{N}\mathbf{1}_{\Delta_{k}(M_{i},M_{j})<0}
\end{align}
where $\mathbf{1}_{X}$ is the indicator function, taking for value
1 if $X$ is true and 0 otherwise, and
\begin{equation}
\Delta_{k}(M_{i},M_{j})=|e_{k}(M_{i})|-|e_{k}(M_{j})|
\end{equation}
Note that, because of the possible presence of ties, one has $\mathrm{SIP}_{i,j}+\mathrm{SIP}_{j,i}\apprle1$.

\subsubsection{Interpretation}

A row of the SIP matrix, provides the SIP values for the corresponding
method over all the other ones. If a new method $M_{1}$ provides
systematic improvement over $M_{2}$, in the sense that it has smaller
absolute errors for all systems in the reference set, one should have
$\mathrm{SIP}_{1,2}=1$. Values smaller than 0.5 indicate a degradation.
Note however that $M_{1}$ can achieve small values of the SIP and
still have better scores (MUE, $Q_{95}$), as a few large improvements
might overwhelm many small degradations. The interest of the SIP indicator
is mainly to alert the user that using a ``better method'' $M_{1}$
can lead to a degradation of results with respect to $M_{2}$, with
a probability close to $(1-\mathrm{SIP}_{1,2})$. 

\subsubsection{Mean SIP}

In order to compare and rank a set of $K$ methods, one defines the
Mean SIP (MSIP) as the mean value of a line of the SIP matrix (excluding
the diagonal)
\begin{equation}
\mathrm{MSIP}(M_{i})=\frac{1}{K}\sum_{j=1}^{K}\mathrm{SIP}_{i,j}\,(1-\delta_{ij})\label{eq:MSIP}
\end{equation}
The largest MSIP value points to a method which in average provides
the best level of improvement over the other methods in the set. Note
that the MSIP is not transferable for comparisons with methods out
of its definition set.

\subsubsection{Representation}

In the same spirit as for correlation matrices, we represent SIP matrices
by a combination of color levels and disks. Here, the color scale
goes from blue (0.0) to red (1.0) with a white midpoint (0.5), and
the area of the disks is proportional to the SIP value. The diagonal
is null. The matrix should be read by row: a row with a majority of
red patches signals a method with good SIP performances. A contrario,
a majority of blue patches on a row indicate a method with poor SIP
performances. The methods are ordered by decreasing value of MSIP. 

\begin{figure}[t]
\noindent \begin{centering}
\includegraphics[width=0.6\textwidth]{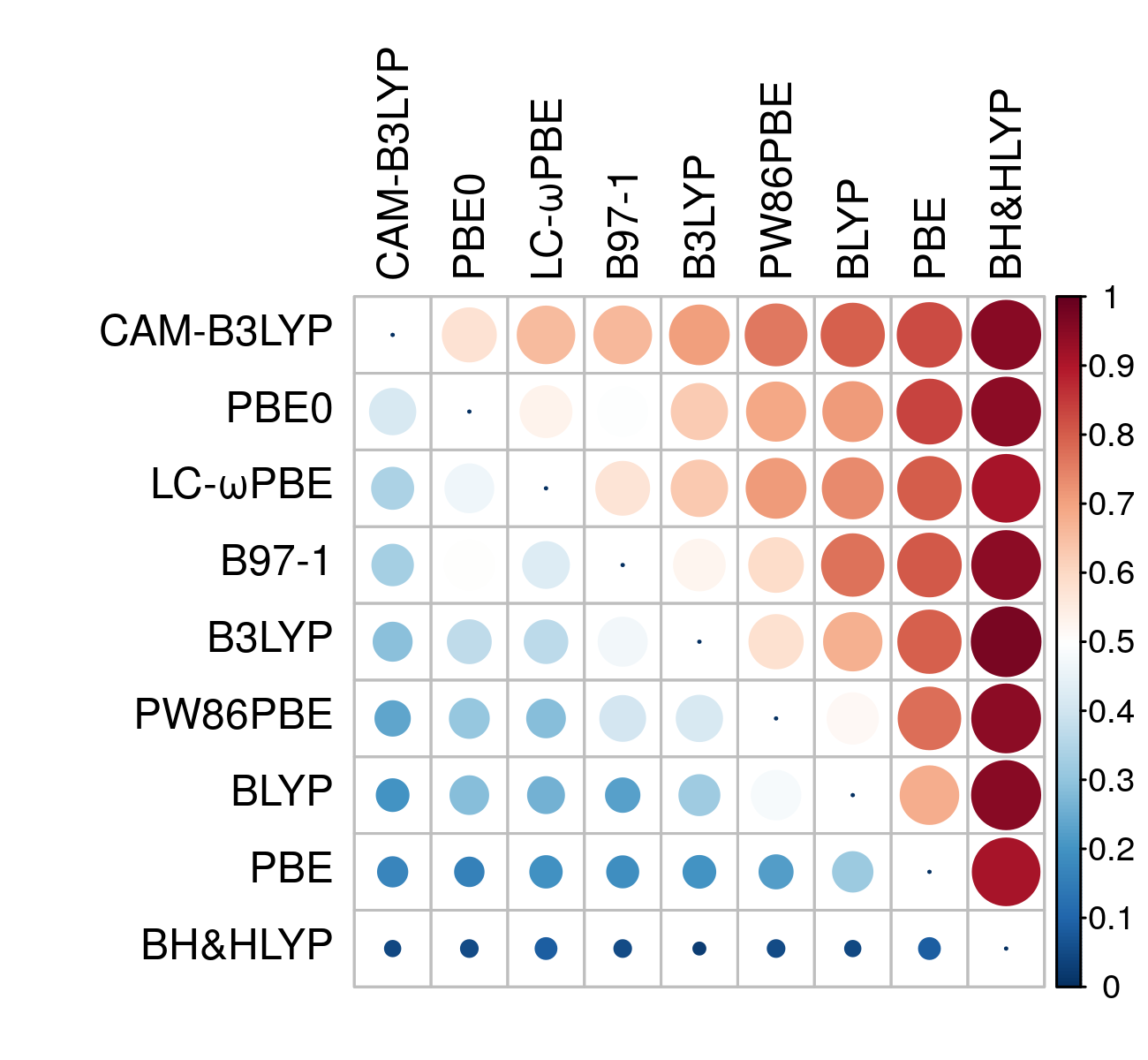}
\par\end{centering}
\noindent \centering{}\caption{\label{fig:SIPMAT-example}SIP matrix for a set of 9 methods compared
on the G99 set of enthalpies (case PER2018, Paper\,II \citep{Pernot2020a}).
The SIP value is color-coded and the area of a disk is proportional
to the corresponding value. A row with a majority of red patches signals
a method with good SIP performances. The methods are ordered by decreasing
value of MSIP (Eq.\,\ref{eq:MSIP}). }
\end{figure}

Fig.\,\ref{fig:SIPMAT-example} provides an example extracted from
a benchmark for intensive atomization energies (case PER2018 in Paper\,II
\citep{Pernot2020a}). It shows clearly that, for this dataset, BH\&HLYP
is problematic, with a row of small blue disks, and is systematically
and strongly outperformed by all other methods. At the opposite, the
row for CAM-B3LYP is the only one to contain exclusively values above
0.5 (reddish disks), albeit CAM-B3LYP does not achieve the best MUE
nor $Q_{95}$ scores within this set of methods \citep{Pernot2018,Pernot2020a}.
This conflict will be further discussed in Paper\,II \citep{Pernot2020a}.

\subsubsection{Mean gain and loss}

In order to appreciate the amplitude of the possible losses or gains
when switching between two methods, we define the mean gain (MG) as
the mean of the negative values of $\Delta_{k}(M_{i},M_{j})$, which
is only defined if $\mathrm{SIP}_{i,j}$ is non-null:
\begin{align}
\mathrm{MG}_{i,j} & =\frac{1}{D_{i,j}}\sum_{k=1}^{N}\mathbf{1}_{\Delta_{k}(M_{i},M_{j})<0}\,\Delta_{k}(M_{i},M_{j})\\
\mathrm{ML}_{i,j} & =-\mathrm{MG}_{j,i}
\end{align}
where by construction the mean loss (ML) is equal the opposite of
the mean gain for the reciprocal comparison. 

These statistics are intended to convey an amplitude of the improvement
of $M_{i}$ over $M_{j}$: MG is therefore a negative value (corresponding
to a decrease of absolute errors), and ML a positive value. Moreover,
the SIP, MG and ML provide a decomposition of the MUE difference between
two methods:
\begin{align}
\Delta_{\mathrm{MUE}_{i,j}} & =\mathrm{MUE}(M_{i})-\mathrm{MUE}(M_{j})\\
 & =\mathrm{SIP}_{i,j}*\mathrm{MG}_{i,j}+\mathrm{SIP}_{j,i}*\mathrm{ML}_{i,j}
\end{align}
This shows that, except for method pairs with extreme SIP values,
any MUE difference is the balance between losses and gains distributed
over the systems. One should not expect that a method with a smaller
MUE will systematically provide better results. 

\subsubsection{ECDF of $\Delta_{k}(M_{i},M_{j})$}

\begin{figure}[t]
\noindent \centering{}\includegraphics[width=0.45\textwidth]{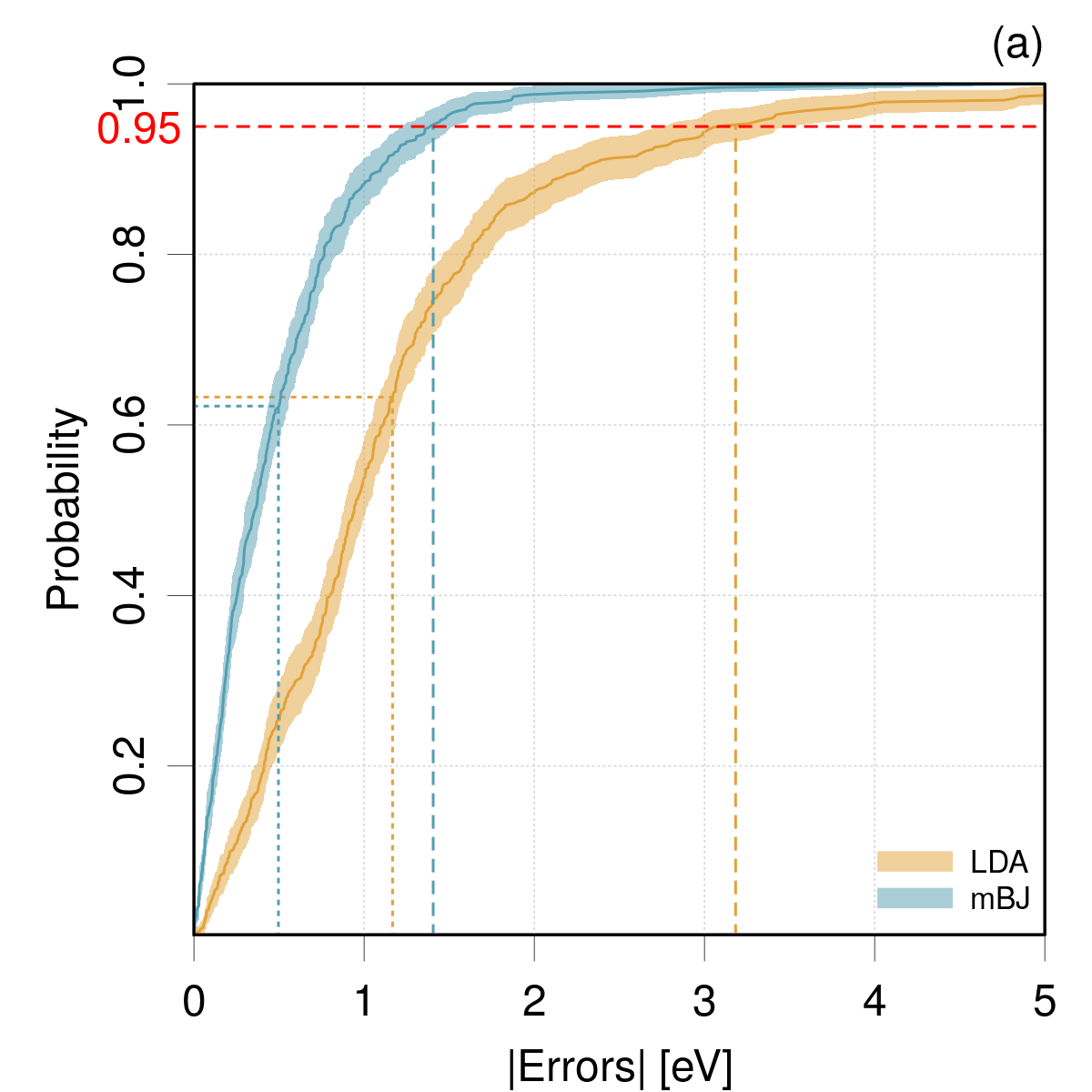}\includegraphics[width=0.45\textwidth]{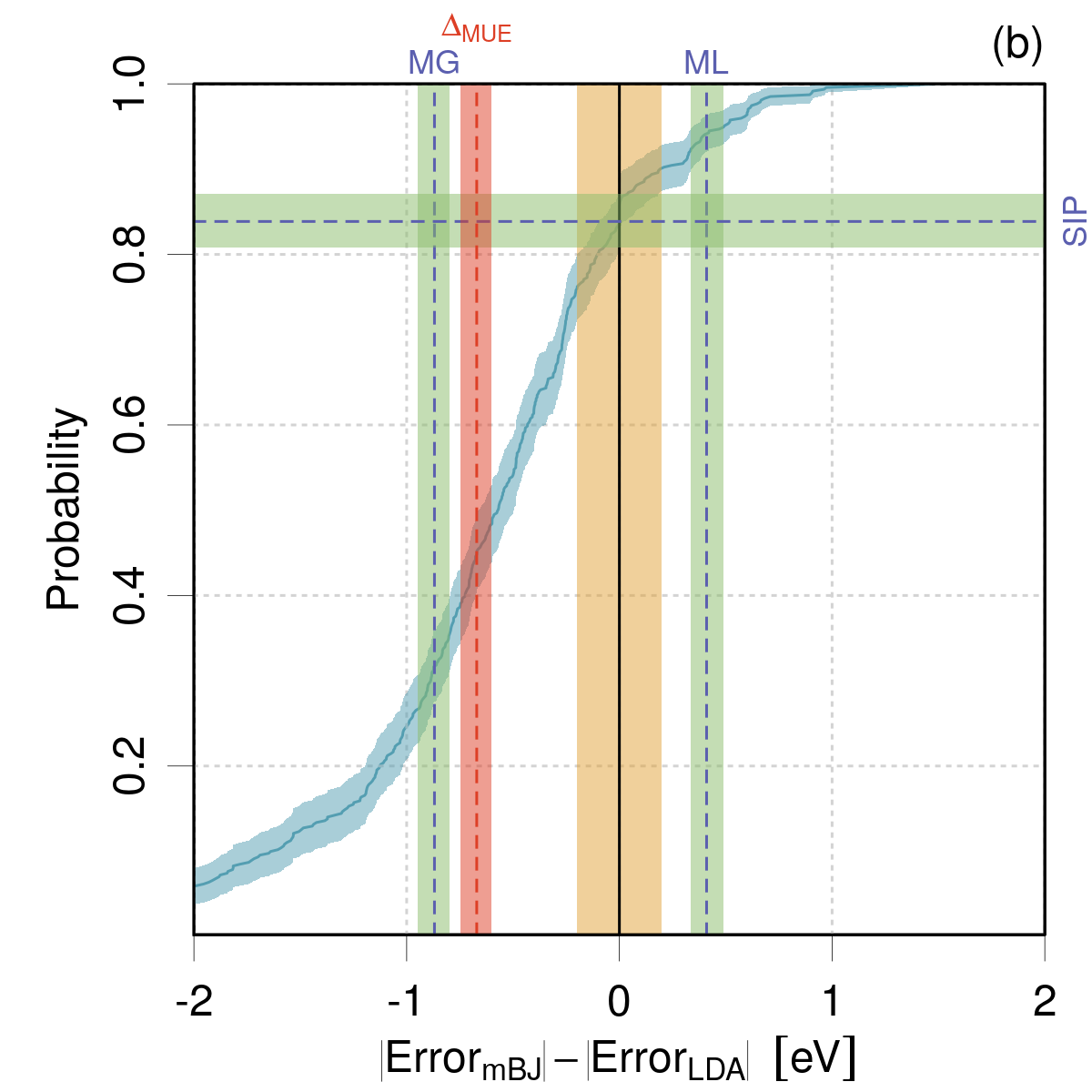}\caption{\label{fig:Delta-example}Statistics of absolute errors on band gaps
for methods mBJ and LDA (case BOR2019, Paper\,II \citep{Pernot2020a})
and of their pair-wise differences: (a) ECDF of two error sets to
be compared. The MUE values are depicted by vertical dotted lines,
and the $Q_{95}$ values by vertical dashed lines. The confidence
bands cover 95\% probability; (b) ECDF of the difference of absolute
errors (blue curve and confidence band). The green- and red-shaded
bands represent 95\,\% confidence intervals for the reported statistics
(SIP: systematic improvement probability; MG: mean gain; ML: mean
loss, $\Delta_{MUE}$: MUE difference). The orange vertical bar represents
an estimated level of uncertainty in the dataset. It is a visual aid
to evaluate the pertinence of the observed differences.}
\end{figure}

The scores (SIP, MG and ML) can be visualized on a single graph of
the Empirical Cumulated Density Function (ECDF) of the differences
of absolute errors between two methods, as shown in Fig.\,\ref{fig:Delta-example}(b).
This example is extracted from the benchmark dataset BOR2019 presented
in Paper\,II \citep{Pernot2020a}, on the prediction of band gaps.
It compares mBJ (MUE = 0.50\,eV) and LDA (MUE = 1.17\,eV). Each
point of the ECDF corresponds to a system of the dataset. Systems
with negative differences are those for which mBJ performs better
than LDA. 

The large MUE difference ($\Delta_{\mathrm{MUE}}$) between these
methods is the balance of a mean gain $\mathrm{MG}=-0.86$\,eV for
85\,\% of the systems (SIP), and a mean loss $\mathrm{ML}=0.37$\,eV
for 15\,\% of the systems. In the hypothesis of a representative
dataset, a user switching from LDA to mBJ has to accept a 15\,\%
risk to see his LDA results degraded in average by 0.37\,eV, and
up to 1\,eV. 

Note that this information is not accessible when considering the
ECDFs of the absolute errors (Fig.\,\ref{fig:Delta-example}(a)).
For the chosen example, the comparison of these ECDFs might leave
the false impression that mBJ has consistently smaller absolute errors
than LDA, which is an artifact due to the missing information about
data pairing (correlation) in this representation. 

\section{Pair-wise comparison of statistics\label{subsec:Pair-wise-comparison-of-1}}

\subsection{The testing framework}

Using the error sets for two methods $M_{1}$ and $M_{2}$, one calculates
the values $s_{1}=S(E_{1})$ and $s_{2}=S(E_{2})$ of a statistic
$S$. A common procedure to compare two values is to test if their
difference is significantly larger than their combined uncertainty,
\emph{i.e.}
\begin{equation}
|s_{1}-s_{2}|>\kappa\thinspace u(s_{1}-s_{2})\label{eq:compare}
\end{equation}
where $u(s_{1}-s_{2})$ is the uncertainty on the difference, and
$\kappa$ is an enlargement factor typically taken as $\kappa=2$
(or 1.96) in metrology \citep{Kacker2010}. In the hypothesis of a
normal distribution for the statistics difference, $\kappa=1.96$
corresponds to a confidence level of 95\,\% for a two-sided test,
implied by the absolute value in Eq.\,\ref{eq:compare}. If one has
evidence that the distribution of differences is not normal, $\kappa$
has to be chosen as the uncertainty enlargement factor providing a
95\,\% confidence interval for this distribution. If the test is
positive, there is less than 5\,\% probability that the difference
between $s_{1}$ and $s_{2}$ is due to sampling effects.

Assuming that $u(s_{1}-s_{2})$ cannot be null nor infinite, it is
convenient to recast the test by using a discrepancy factor
\begin{equation}
\xi(s_{1},s_{2})=\frac{|s_{1}-s_{2}|}{u(s_{1}-s_{2})}\label{eq:discFac-1}
\end{equation}
to be compared to the threshold $\kappa$. A probability value ($p$-value)
corresponding to $\xi$ is derived from the cumulated density function
of the expected distribution for $\xi$. For instance
\begin{align}
p_{t} & =1-\Phi_{H}(\xi)\label{eq:pt-1}\\
 & =2*\left(1-\Phi(\xi)\right)\label{eq:pt}
\end{align}
where $\Phi_{H}(.)$ is the cumulative distribution function (CDF)
of the standard half-normal distribution \citep{Leone1961}, and $\Phi(.)$
is the CDF of the standard normal distribution. The half-normal distribution
is used to account for the absolute value in Eq.\,\ref{eq:discFac-1}.
The $t$ index of $p_{t}$ refers here to the analogy with the two-sample
$t$-test for equal means \citep{Snedecor1989}. $p_{t}$ is the probability
to obtain values of $\xi$ equal to or larger than the calculated
value, assuming that the null hypothesis, $S(E_{1})=S(E_{2})$, is
true. For testing, one chooses a probability threshold corresponding
to $P(\xi>\kappa=1.96)=0.05$ . For $p_{t}$ above this value, one
chooses not to reject the hypothesis that the observed difference
between $s_{1}$ and $s_{2}$ is due to random effects. 

In order to be able to estimate $p_{t}$, one needs to evaluate the
uncertainty on the difference of $s_{1}$ and $s_{2}$. Formally,
it can be obtained by the combination of variances \citep{GUM}
\begin{equation}
u(s_{1}-s_{2})=\sqrt{u^{2}(s_{1})+u^{2}(s_{2})-2\mathrm{cov}(s_{1},s_{2})}\label{eq:u-diff-stat}
\end{equation}
The usefulness of this formula depends on several assumptions (theoretical
limits of the statistics not within a high probability interval around
their values, symmetry of error intervals... \citep{Nicholls2014,Nicholls2016}).
Nevertheless, it shows that the covariance between statistics can
have a major effect on the amplitude of $u(s_{1}-s_{2})$. In the
limit of very strong positive correlation, the uncertainty on the
difference can become very small, impacting $\xi(s_{1},s_{2})$ and
$p_{t}$.

To estimate the effect of correlation on the comparison of scores,
we introduce a variant $p_{unc}$ (uncorrelated) of $p_{t}$, based
on a version of the discrepancy ignoring correlation
\begin{align}
\xi_{unc}(s_{1},s_{2}) & =\frac{|s_{1}-s_{2}|}{\sqrt{u(s_{1})^{2}+u(s_{2})^{2}}}\label{eq:xiUnc}\\
p_{unc} & =2*\left(1-\Phi(\xi_{unc})\right)\label{eq:Punc}
\end{align}
In the hypothesis of  mostly positive correlations for the statistics
of interest (MUE and $Q_{95}$; Appendix\,\ref{sec:Covariance-of-scores}),
$p_{unc}$ is expected to overestimate $p_{t}$.

\subsection{Bootstrap-based comparison of statistics\label{subsec:Bootstrap-based-comparison-of}}

Several strategies can be considered to compare pairs of statistics
$(s_{1},s_{2})$ through a $p$-value.

\subsubsection{Estimate $u(s_{1})$, $u(s_{2})$ and $\mathrm{cov}(s_{1},s_{2})$}

The uncertainty on the statistics of interest (except for the MSE
and RMSD) and their covariance are not, to our knowledge, available
in analytical form. In consequence, one has to use a numerical procedure,
such as the bootstrap to estimate them \citep{Efron1979,Hesterberg2015}.
The application of the bootstrap to individual terms of Eq.\,\ref{eq:u-diff-stat}
will result in an accumulation of statistical uncertainties. Besides,
the estimation of covariances is known to be very sensitive to outliers.
This approach is clearly suboptimal and is not recommended.

\subsubsection{Estimate directly $u(s_{1}-s_{2})$ }

A better approach in the present context is to estimate directly (by
bootstrap) the uncertainty on the difference of scores. This relieves
underlying hypotheses in Eq.\,\ref{eq:u-diff-stat}, and enables
the explicit correlation of samples of $s_{1}$ and $s_{2}$ through
paired-data sampling. However, estimating a discrepancy factor leads
us to use Eq.\ref{eq:pt} to estimate the $p$-value, with the associated
normality hypothesis. 

\subsubsection{Generalized $p$-value}

The use of the generalized $p$-value ($p_{g}$), as proposed by Wilcox
and Erceg-Hurn \citep{Liu1997,Wilcox2012} (method M; \emph{cf.} Algorithm\,\ref{alg:methodM}),
conveniently avoids to estimate $u(s_{1}-s_{2})$, and the incurring
normality hypothesis of $p_{t}$. It is based on a simple counting
of null and negative bootstrapped differences of statistics with paired
samples. If $S(E_{1})=S(E_{2})$, one expects that the bootstrap sample
will generate positive and negative values of their difference in
equal amounts. In this case, $p^{*}\simeq1-p^{*}\simeq0.5$ and $p_{g}$
is close to 1. Note that the null values in the differences sample
are shared equally between the positive and negative values. On the
opposite, if there is a small proportion $p^{*}$ of negative values,
the mean of the differences sample should be positive, different from
zero. The smaller $p^{*}$ the farther the mean from zero, and the
lower the probability of the null, $S(E_{1})=S(E_{2})$, hypothesis.
The symmetric case occurs for large values of $p^{*}$ (small values
of $1-p^{*}$). As the sign of the difference is irrelevant, a factor
two is applied to estimate $p_{g}$. The identity of this algorithm
with the analytical $p$-value for the comparison of the means of
normal samples is established in Appendix\,\ref{subsec:Estimation-of--values}.
\begin{algorithm}
Input: Two paired error sets $E_{1}$, $E_{2}$ of size $N$, a statistic
estimator $S$, and a number of bootstrap samples $B$
\begin{enumerate}
\item Bootstrap the statistics difference
\begin{enumerate}
\item For $j=1:B$
\begin{enumerate}
\item Generate a $N$-sample of paired data with replacement $\longrightarrow\left(E_{1}^{*},E_{2}^{*}\right)$ 
\item Estimate $d_{j}=S(E_{1}^{*})-S(E_{2}^{*})$ 
\end{enumerate}
\end{enumerate}
\item Calculate a generalized $p$-value to test $S(E_{1})=S(E_{2})$

$p_{g}=2\min(p^{*},1-p^{*})$, where\\
 $p^{*}=(A+0.5C)/B$\\
$A=\sum_{i=1}^{B}1_{d_{i}<0}$ \\
 $C=\sum_{i=1}^{B}1_{d_{i}=0}$ 
\end{enumerate}
\caption{\label{alg:methodM}Method M: testing the equality of a statistic
$S$ for two paired samples by bootstrap and a generalized $p$-value
($p_{g}$) \citep{Wilcox2012}.}
\end{algorithm}

The use of paired samples is essential to capture inter-statistics
correlations. Wilcox and Erceg-Hurn \citep{Wilcox2012} have shown
that their method M provides a well controlled level of type I errors
(false positive) for the comparison of quantiles at the 0.05 level.
They estimated that dataset sizes of $N\ge30$ are necessary when
comparing quantiles up to 0.9. This applies to the MUE, which we have
shown to lie typically between the 0.5 and 0.75 quantiles \citep{Pernot2018}.
Using the same protocol, we estimated that for the comparison of $Q_{95}$
values at the same 0.05 level, $N\ge60$ is requested. Details are
presented in Appendix\,\ref{sec:Type-I-error}. 

\subsection{Rank inversion probability $P_{inv}$\label{subsec:Rank-inversion-probability}}

In a previous article \citep{Pernot2018}, we defined a ranking inversion
probability 
\begin{equation}
P_{inv}=P(S_{1}<S_{2}|s_{1}>s_{2})\label{eq:defPinv}
\end{equation}
and estimated it using the hypothesis of a normal distribution for
the difference of statistics. Using Equations\,\ref{eq:xiUnc}-\ref{eq:Punc},
this former estimation can be reformulated as 
\begin{align}
P_{inv} & =\Phi(0;\mu=s_{1}-s_{2},\sigma=\sqrt{u^{2}(s_{1})+u^{2}(s_{2})})\\
 & =\Phi(0;\mu=\xi_{unc})\\
 & =\Phi(-\xi_{unc})\\
 & =1-\Phi(\xi_{unc})\\
 & =p_{unc}\thinspace/\thinspace2
\end{align}
where the unspecified parameters of the normal cumulative distribution
function $\Phi(x;\mu,\sigma)$ are their standard values ($\mu=0$,
$\sigma=1$). The link to $p_{unc}$ shows the limitations of our
previous estimation of $P_{inv}$, \emph{i.e.}, the normality hypothesis
and the neglect of error sets correlations. 

Using the same difference statistics used for $p_{g}$ (Algorithm\,\ref{alg:methodM}),
one can generalize Eq.\,\ref{eq:defPinv} by defining $P_{inv}$
as the probability to have differences in the bootstrap sample with
a sign opposite to the reference one ($\mathrm{sign}(s_{1}-s_{2})$)
\begin{align}
P_{inv} & =\frac{1}{B}\left(\sum_{i=1}^{B}1_{\mathrm{sign}(d_{i})\ne\mathrm{sign}(s_{1}-s_{2})}-\sum_{i=1}^{B}1_{d_{i}=0}\right)\label{eq:pinv-new}
\end{align}
where $B$ is the number of bootstrap samples and the null differences
(with sign 0) are compensated for. Enforcing the condition $s_{1}>s_{2}$
in Eq.\,\ref{eq:defPinv}, one gets $\mathrm{sign}(s_{1}-s_{2})=1$,
and finally
\begin{align}
P_{inv} & =\frac{1}{B}\left(\sum_{i=1}^{B}1_{\mathrm{sign}(d_{i})\ne1}-\sum_{i=1}^{B}1_{d_{i}=0}\right)\\
 & =\frac{1}{B}\left(\sum_{i=1}^{B}1_{d_{i}\le0}-\sum_{i=1}^{B}1_{d_{i}=0}\right)\\
 & =\frac{1}{B}\sum_{i=1}^{B}1_{d_{i}<0}\\
 & \simeq p_{g}\thinspace/\thinspace2\label{eq:pinv-vs-pg}
\end{align}
where the relation to $p_{g}$ (Algorithm\,\ref{alg:methodM}) assumes
a negligible probability to have null statistics differences and exploits
the fact that $\sum_{i=1}^{B}1_{d_{i}<0}<\sum_{i=1}^{B}1_{d_{i}>0}$
if $s_{1}>s_{2}$. 

\subsection{Ranking probability matrix $\mathbf{P}_{r}$\label{subsec:Ranking-probability-matrix}}

A measure of the reliability of a statistic-based ranking can be estimated
by bootstrap \citep{Hall2009}. This approach has notably been used
by Proppe and Reiher \citep{Proppe2017} to study how the sample size
affects the probability for a DFA to be ranked at first place on the
basis of its prediction uncertainty. We apply it here to compute,
for a set of $K$ methods scored by a statistic $S$, a ranking probability
matrix $\mathbf{P}_{r}$ giving, for each method, its probability
to have any rank
\begin{equation}
P_{r,jk}=P(\mathrm{rank}(S_{j})=k);\thinspace j,k=1,\ldots,K
\end{equation}
The algorithm to generate this matrix is described in Algorithm\,\ref{alg:bs-rank}.
\begin{algorithm}
Input: $K$ paired error sets, $E_{1},\ldots,E_{K}$ of size $N$,
a statistic estimator $S$, and a number of bootstrap samples $B$
\begin{enumerate}
\item Bootstrap the ranks
\begin{enumerate}
\item For $j=1:B$
\begin{enumerate}
\item Generate a $N$-sample of paired data with replacement $\longrightarrow\left(E_{1}^{*},\ldots,E_{K}^{*}\right)$ 
\item Estimate the statistics vector $S^{*}=\left(S(E_{1}^{*}),\ldots,S(E_{K}^{*})\right)$
\item Estimate the ranks by increasing order of $S^{*}$: $O_{j}^{*}=\mathrm{order}(S^{*})$,
\\
where $O_{j}^{*}$ is a $K$-vector of integer values. 
\end{enumerate}
\end{enumerate}
\item Estimate for each method its probability to have any rank

\[
P_{r,jk}=\frac{1}{B}\sum_{i=1}^{B}1_{O_{ij}^{*}=k}
\]

\end{enumerate}
\caption{\label{alg:bs-rank}Estimating the rank probabilities for a set of
methods.}
\end{algorithm}

\subsubsection{Representations}

\begin{figure}[t]
\noindent \begin{centering}
\includegraphics[viewport=300bp 400bp 1800bp 1700bp,clip,height=7cm]{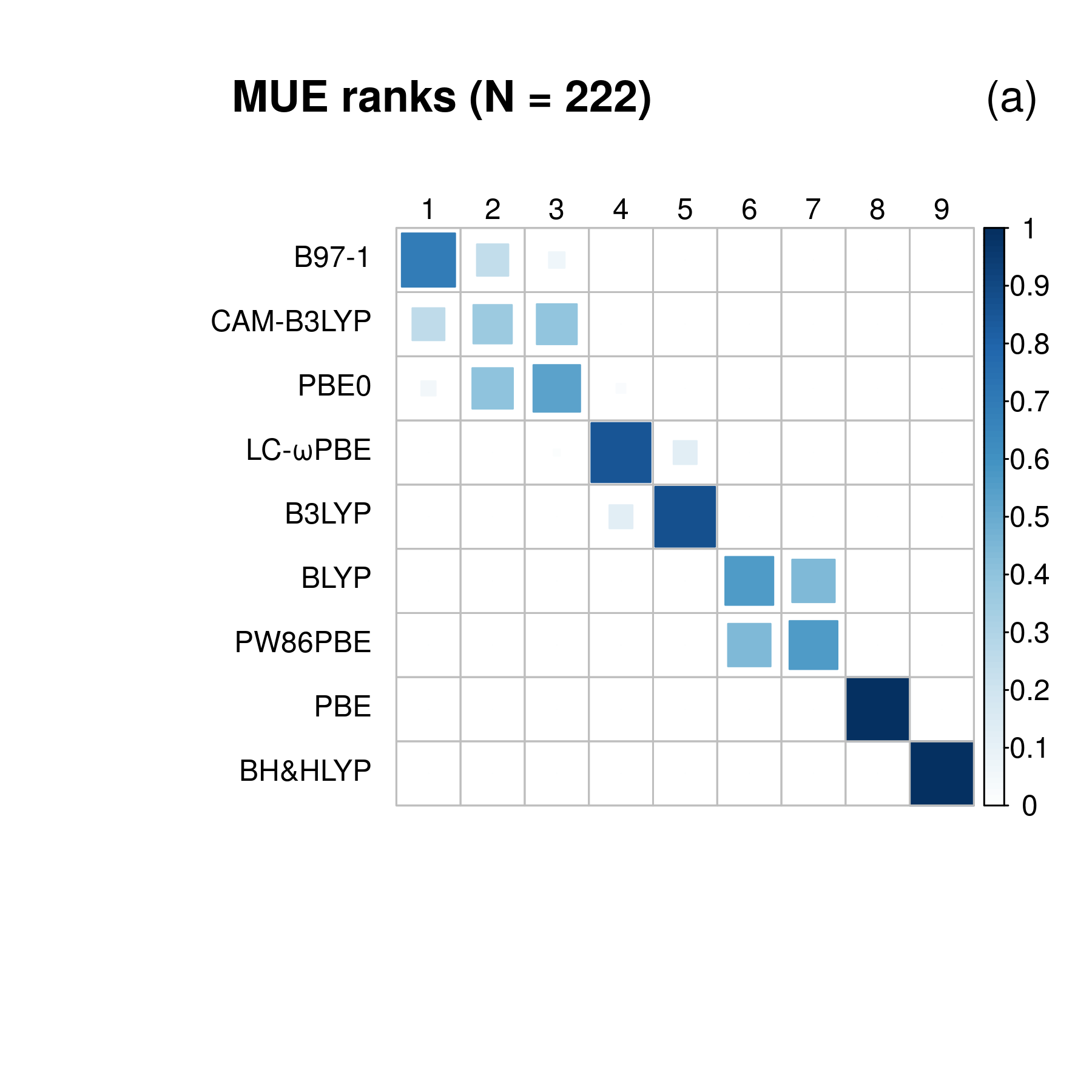}\includegraphics[viewport=0bp 170bp 1800bp 1800bp,clip,height=7cm]{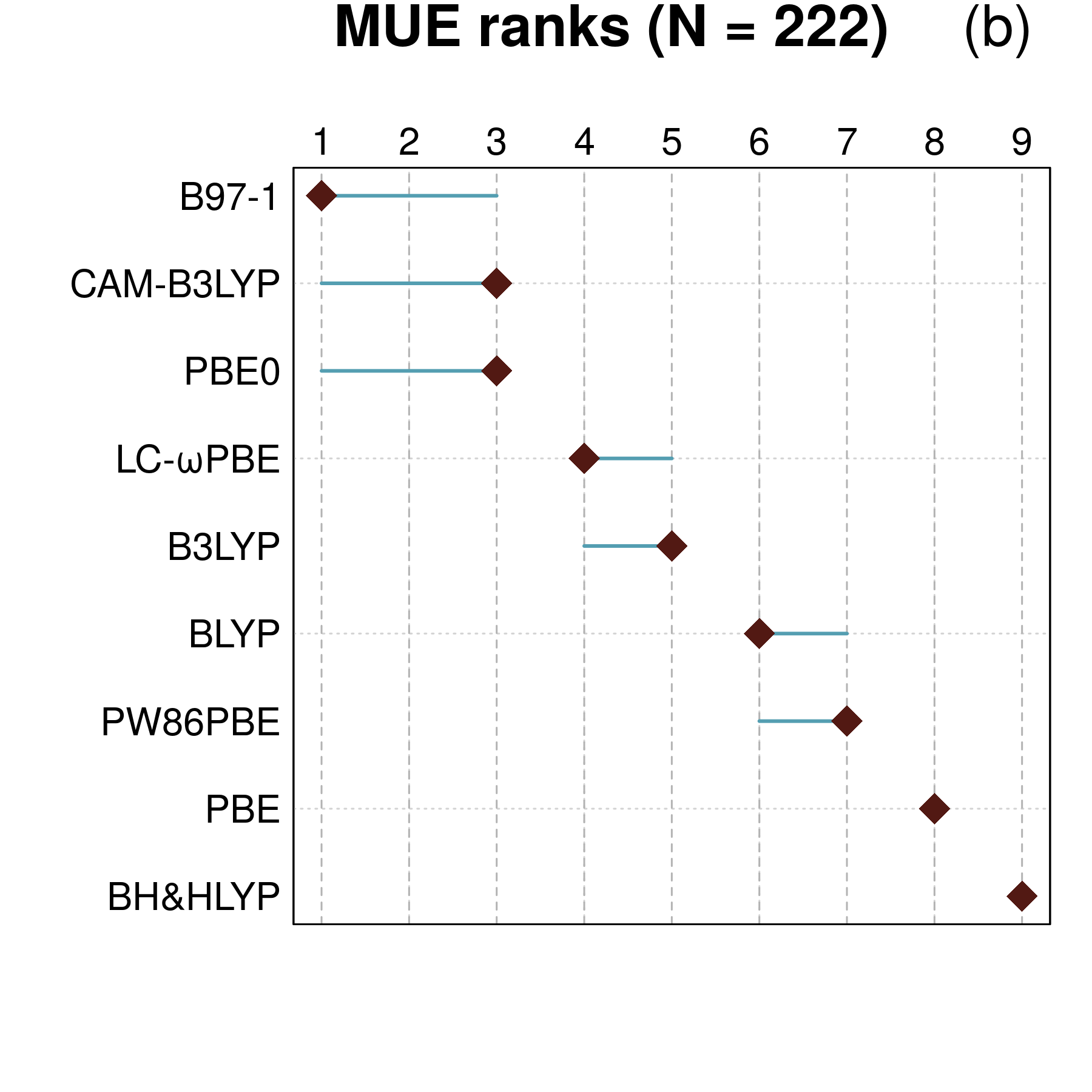}
\par\end{centering}
\noindent \centering{}\caption{\label{fig:bsRank}Graphical representations of a MUE ranking probability
matrix $\mathbf{P}_{r}$ : (left) color levels image of the ranking
probability matrix; (right) summary of the ranking probability matrix
by the modes (diamonds) and 90\,\% probability intervals. The data
are taken from the case PER2018 (\emph{cf}. Paper\,II \citep{Pernot2020a}).
Both representations indicate a possible ranking inversion between
B97-1, CAM-B3LYP and PBE0, \emph{i.e.}, the reference ranking based
on the MUE is not certain for this trio. Similar problems occur within
two other groups, notably BLYP and PW86PBE. The ranks of PBE (8) and
BH\&HLYP (9) are well established. }
\end{figure}

Two representations for this matrix are proposed by Hall and Miller
\citep{Hall2009}, either a combined color-levels\,/\,symbol-size
image (Fig.\,\ref{fig:bsRank}(a)), or a summary by mode and probability
intervals (Fig.\,\ref{fig:bsRank}(b)). In the following, we will
use mostly the levels image representation which we find easier to
read and interpret.\footnote{A summary in results tables can also be considered, by reporting for
each method its mode in ranking probability and the corresponding
probability, which indicates the strength of this rank. } 

\subsubsection{Remarks}
\begin{itemize}
\item As discussed by Hall and Miller \citep{Hall2009}, the standard bootstrap
($N$-out\,of-$N$ sampling) tends to underestimate the dispersion
of the ranks. Better estimates would be obtained by a $N'$-out\,of-$N$
sampling ($N'<N$), but the best choice of $N'$ is problem-dependent
and is left to the appreciation of the analyst. For the sake of simplicity,
and until further guidance on the optimal choice of $N'$, we consider
here that the standard method provides a reasonable qualitative appreciation
of ranking uncertainties. An example with $N'=N/3$ is presented in
case DAS2019 of Paper\,II \citep{Pernot2020a}.
\item As a general trend, one expects that ranking uncertainty will increase
for smaller error sets, but might also increase with the number $K$
of compared methods, notably if several methods have similar performances.
 
\end{itemize}

\section{Implementation\label{subsec:Implementation}}

Calculations have been made in the\texttt{ R} language \citep{CiteR},
using several packages, notably for the bootstrap\texttt{ (boot} \citep{R-boot}).
Bootstrap estimates are based on 1000 replicates. 

\paragraph{Quantiles.}

Wilcox and Erceg-Hurn \citep{Wilcox2012} recommend the use of the
Harrell and Davis method for quantiles estimation \citep{Harrell1982},
which provides a better stability for the bootstrap sampling of quantiles.
The relevance of this choice is illustrated in Appendix\,\ref{sec:Simulated-example}.
In the case studies of Paper\,II \citep{Pernot2020a}, all quantiles
are estimated by the Harrell and Davis method \citep{Harrell1982},
as implemented in package \texttt{WSR2} \citep{Wilcox2012,Wilcox2018,R-WRS2}. 

\paragraph{Correlation.}

The estimation of correlation coefficients by the standard Pearson
method is reputed to be very sensitive to the presence of outliers
\citep{Wilcox2018}. As the presence of a small amount of outliers
is a frequent feature of the benchmarking data sets, we use the more
robust rank-correlation (Spearman) method, unless otherwise specified. 

\paragraph{Code.}

The application \texttt{ErrView} implementing the methods described
in this article (and more) and the corresponding datasets are archived
at \url{https://github.com/ppernot/ErrView} (DOI: 10.5281/zenodo.3628489);
a test web interface is also freely accessible at \url{http://upsa.shinyapps.io/ErrView}.

\section{Conclusions\label{sec:Conclusions}}

In this article, we proposed several tools to test the robustness
of rankings or comparisons of methods based on error statistics for
non-exhaustive, limited size datasets. In order to avoid hypotheses
on the errors distributions, bootstrap-based methods were adopted,
as suggested by Proppe and Reiher \citep{Proppe2017} for the estimation
of prediction uncertainty of DFT methods. Special care has been taken
to use (robust) methods best adapted to provide reliable results for
small datasets. 

We introduced the systematic improvement probability (SIP) which is
independent of other descriptive statistics. We have shown that the
use of MUE for ranking hides a complex interplay between genuine method
improvements and error cancellations inherent to most computational
chemistry methods. In particular, we have shown how a difference in
MUE is a balance between gains and losses in absolute errors. Estimation
of the systematic improvement probability (SIP), the mean gain (MG)
and mean loss (ML) statistics can help understand this balance, and
to assess the risks for a user of switching between two methods.

When considering pairs of methods, we generalized our previous definition
of the inversion probability $P_{inv}$ to account for correlations
between statistics and relieve a normal distribution hypothesis. The
link of $P_{inv}$ to $p$-values for the comparison of two values
of a statistic has been established. 

Finally, the ranking probability matrix $\mathbf{P}_{r}$ for a chosen
statistic provides a clear diagnostic on the robustness of the corresponding
ranking. 

All these tools are put to test in Paper\,II \citep{Pernot2020a},on
nine datasets from the recent benchmark literature. 

\section*{Supplementary Information}

The data that support the findings of this study are openly available
in Zenodo at\\
 \href{http://doi.org/10.5281/zenodo.3678481}{http://doi.org/10.5281/zenodo.3678481}\citep{SIPdata2020}.

\appendix
\appendixpage

\section{Estimation of the mean value and its uncertainty\label{sec:Estimation-of-the}}

Let us consider the mean (signed) value of the errors (MSE). In absence
of uncertainty, it is defined as
\begin{equation}
\overline{e}=\frac{1}{N}\sum_{i=1}^{N}e_{i}
\end{equation}
and its uncertainty (standard error) is estimated as
\begin{equation}
u(\overline{e})=\sqrt{\frac{s_{e}^{2}}{N}}\label{eq:uref}
\end{equation}
where $s_{e}^{2}$ is a sample-based estimator of the population variance
\begin{equation}
s_{e}^{2}=\frac{1}{N-1}\sum_{i=1}^{N}(e_{i}-\overline{e})^{2}
\end{equation}
Eq.\,\ref{eq:uref} gives the well-known dependence of the MSE uncertainty
with the dataset size for independent and identically distributed
(\emph{i.i.d.) }errors, assuming a finite variance, which might exclude
error sets with heavy-tailed distributions, \emph{e.g.}, Cauchy. \footnote{Note that $u(\overline{e})$ in Eq.\,\ref{eq:uref} does not account
for the uncertainty on $s_{e}$. Taking this factor into account leads
to a larger uncertainty, which can be estimated as $u(\overline{e})=\sqrt{(N-1)/(N-3)}\thinspace s_{e}/\sqrt{N}$
\citep{Kacker2003}. This formula is based on the properties of the
Student's-\emph{t} distribution \citep{Evans2000}. The impact of
the correction factor is notable only for very small datasets (smaller
than 3\,\% for $N\ge30$), and we will consider the standard formula
.}

If uncertainty on errors $u(e_{i})$ is negligible, $s_{e}$ is an
estimation of the standard deviation of the errors distribution $\sigma$,
which represents the dispersion of model errors. If the reference
data are uncertain, $s_{e}$ quantifies a dispersion due to both model
errors and reference data uncertainty. In consequence, it overestimates
the dispersion of model errors, and specific models have to be designed
if one wishes to estimate this specific contribution \citep{Pernot2015,Proppe2017}.
This points to the necessity of using accurate reference data if the
benchmark based on standard statistics is to reflect the properties
of the studied methods.

To be more specific, in the presence of uncertainty on errors, the
weighted mean is the maximum likelihood estimator of the distribution
mean under normality assumptions \citep{Bevington1992}
\begin{align}
\overline{e} & =\sum_{i=1}^{N}w_{i}e_{i}\\
w_{i} & =\frac{u(e_{i})^{-2}}{\sum_{j=1}^{N}u(e_{j})^{-2}}\label{eq:wRefUnc}
\end{align}
giving less weight to the more uncertain data. Direct application
of the combination of variances to this expression leads to \citep{Bevington1992}
\begin{equation}
u(\overline{e})^{2}=\frac{1}{\sum_{j=1}^{N}u(e_{j})^{-2}}
\end{equation}
Note that in the case of identical uncertainty for all data, one recovers
the expression for the unweighted case (Eq.\,\ref{eq:uref}). 

The validity of this estimation has to be tested by computing the
weighted chi-squared
\begin{equation}
\chi_{w}^{2}=\sum_{i}\frac{(e_{i}-\overline{e})^{2}}{u(e_{i})^{2}}\label{eq:Birge}
\end{equation}
If the errors on the reference data are assumed to be normally distributed,
$\chi_{w}^{2}$ has a chi-squared distribution with $N-1$ degrees
of freedom ($\chi_{N-1}^{2}$). $\chi_{w}^{2}$ should be close to
the mean of this distribution, $N-1$, and lie within its 95\,\%
high probability interval. If $\chi_{w}^{2}$ is too small, the $u(e_{i})$
are over-estimated and should be reconsidered, or the benchmarked
method is over-fitting the data, which is unlikely, unless the method
is parametric and has been calibrated on this same dataset. If $\chi_{w}^{2}$
is too large, there is an excess of variance in the $E_{M}$ error
set \citep{Kacker2004,Rukhin2009,Rivier2014}. In the typical benchmarking
of computational chemistry methods, this is generally the case because
of the extraneous dispersion due to model errors. To ensure the statistical
validity of the weighted mean and its uncertainty, one has therefore
to define a more complex error model, considering explicitly the two
sources of dispersion, and to redefine the weights, accounting for
the excess of variance and possible biases in the error sets \citep{Lejaeghere2014,Lejaeghere2014a,Pernot2015,DeWaele2016,Proppe2017}. 

If one stipulates that the dispersion of the errors is the combined
effect of model error and reference data uncertainty, one can redefine
the weights as \citep{Rukhin2009}
\begin{equation}
w_{i}=\frac{\left(\sigma^{2}+u(e_{i})^{2}\right)^{-1}}{\sum_{j=1}^{N}\left(\sigma^{2}+u(e_{j})^{2}\right)^{-1}}\label{eq:weights-IRWLS}
\end{equation}
where $\sigma^{2}$ is the variance of model errors. With these new
weights, 
\begin{equation}
u(\overline{e})^{2}=\frac{1}{\sum_{j=1}^{N}\left(\sigma^{2}+u(e_{j})^{2}\right)^{-1}}\label{eq:uwmean}
\end{equation}
converges properly to the standard limit when the reference data errors
become negligible before the model errors. The model error variance
$\sigma^{2}$ can be estimated by decomposing the total variance of
the errors into the variance of model errors plus the mean variance
of the data (known as Cochran's ANOVA estimate \citep{Kacker2004,Rivier2014})
\begin{equation}
\mathrm{var}(e)=\sigma^{2}+\frac{1}{N}\sum_{j=1}^{N}u(e_{j})^{2}\label{eq:dispmod}
\end{equation}
This variance analysis ensures that $\chi_{w}^{2}$ is correct. Note
that other reweighting schemes exist \citep{Kacker2004,Rivier2014},
but Cochran's is the simplest. Besides, reweighting methods are iterative:
$\sigma$ depends on $\overline{e}$, which itself depends on $\sigma$.

If the dispersion of reference data uncertainties is small, \emph{i.e.},
smaller than the model errors contribution, one can reasonably consider
that the weights are identical and that the unweighted mean can be
used. Formally, its uncertainty (Eq.\,\ref{eq:uwmean}) depends on
$\sigma$, which can be directly estimated through Eq.\,\ref{eq:dispmod},
but by construction, one will recover results given by Eq.\,\ref{eq:uref}. 

One will therefore consider that, unless a large dispersion of reference
data uncertainty is observed, these uncertainties can be ignored in
the estimation of the mean and its standard error. Otherwise, one
should use the weighted mean with the standard uncertainty estimate.\footnote{Note that the dispersion of model errors $\sigma$ is related to the
model prediction uncertainty and is a score of interest for the ranking
of models \citep{Pernot2015,Pernot2018}. } 

An advanced modeling of uncertainty sources is crucial if one wishes
a reliable estimate of the MSE, and of the various uncertainty contributions
\citep{Pernot2015}. In standard benchmarking, the aim is mostly to
compare methods, knowing that the reference datasets are incomplete.
If reference data uncertainty plays a significant role -- that would
be the case if data with very different uncertainty levels were aggregated
in the dataset -- one might assume that its impact will be the same
for all methods to be compared. The values of the dispersion statistics
will be consistently overestimated for all methods. As long as one
is not interested in the accurate estimation of the underlying properties
of the error distributions, such as the model prediction uncertainty
\citep{Pernot2015,Proppe2017}, it is simpler to rely on unweighted
schemes and properly curated datasets.

\section{Numerical study of the correlation of nonlinear statistics\label{sec:Covariance-of-scores}}

To illustrate the transfer of correlation from errors sets $E_{1}$
and $E_{2}$ to their statistics, one assumes that they are described
by a bivariate distribution with prescribed correlation coefficient
$\rho$. From this distribution, one generates random samples $E_{1}^{*}$
and $E_{2}^{*}$ and one estimates the statistics values $s_{1}^{*}=S(E_{1}^{*})$
and $s_{2}^{*}=S(E_{2}^{*})$. $\mathrm{cor}(s_{1},s_{2})$ is finally
estimated from $s_{1}^{*}$ and $s_{2}^{*}$ samples. 

The error sets correlation coefficient $\rho$ is varied between -1
and 1, and the resulting correlation coefficients are estimated for
the MSE, MUE and $Q_{95}$ statistics. The dataset size is $N=100$
and Monte Carlo samples size is $M=10^{3}$ . 

\begin{figure}[!tb]
\noindent \begin{centering}
\includegraphics[width=0.32\textwidth]{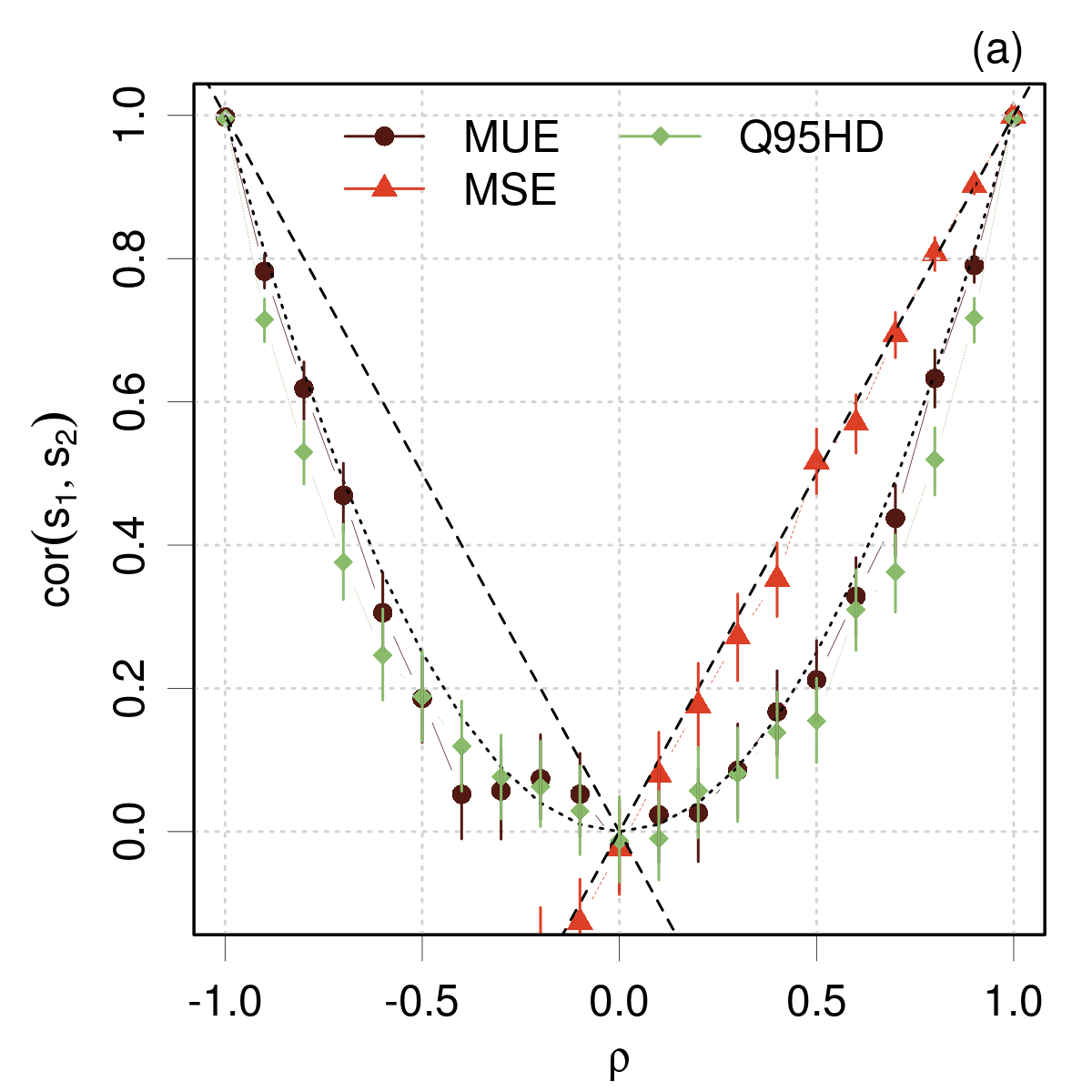}\includegraphics[width=0.32\textwidth]{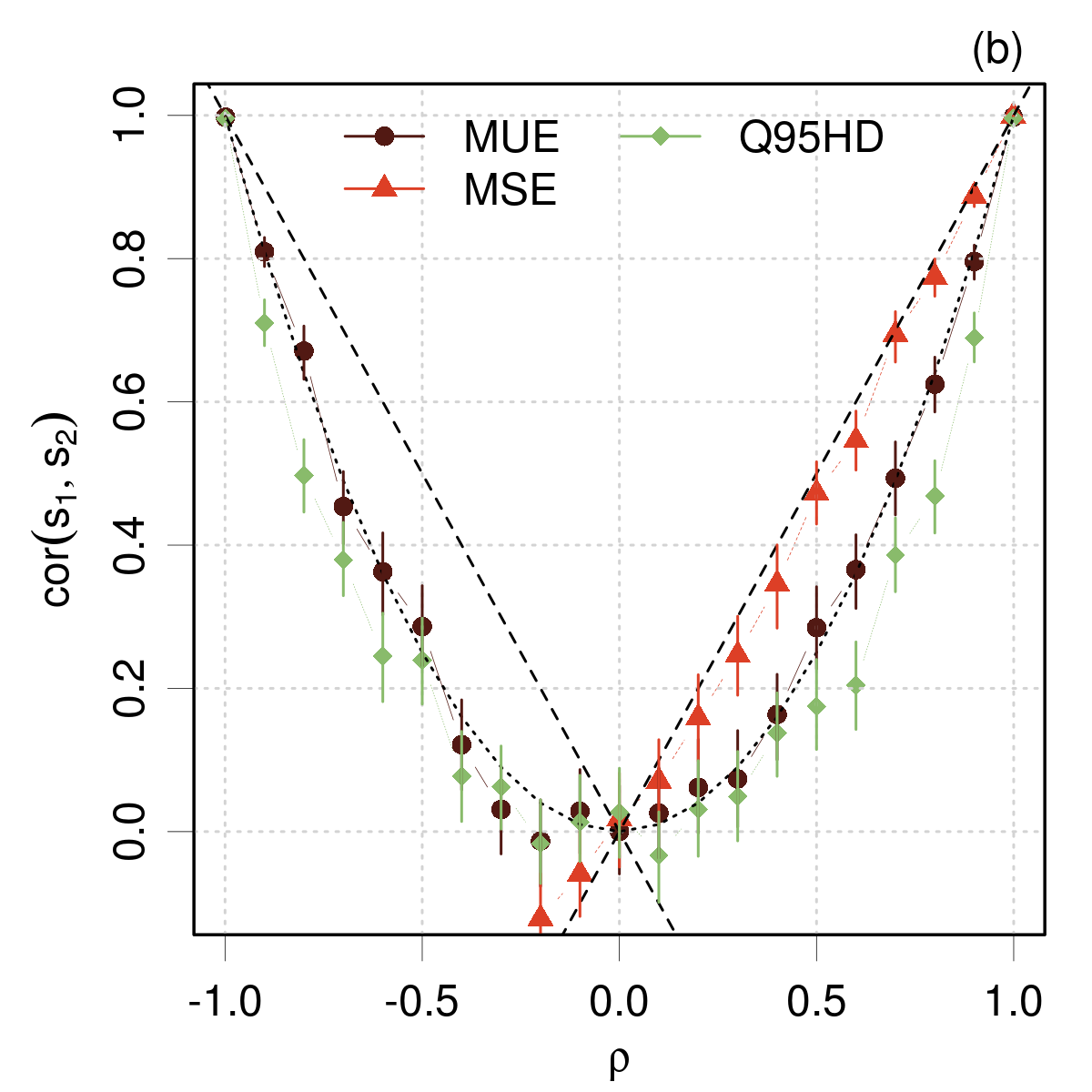}
\par\end{centering}
\noindent \begin{centering}
\includegraphics[width=0.32\textwidth]{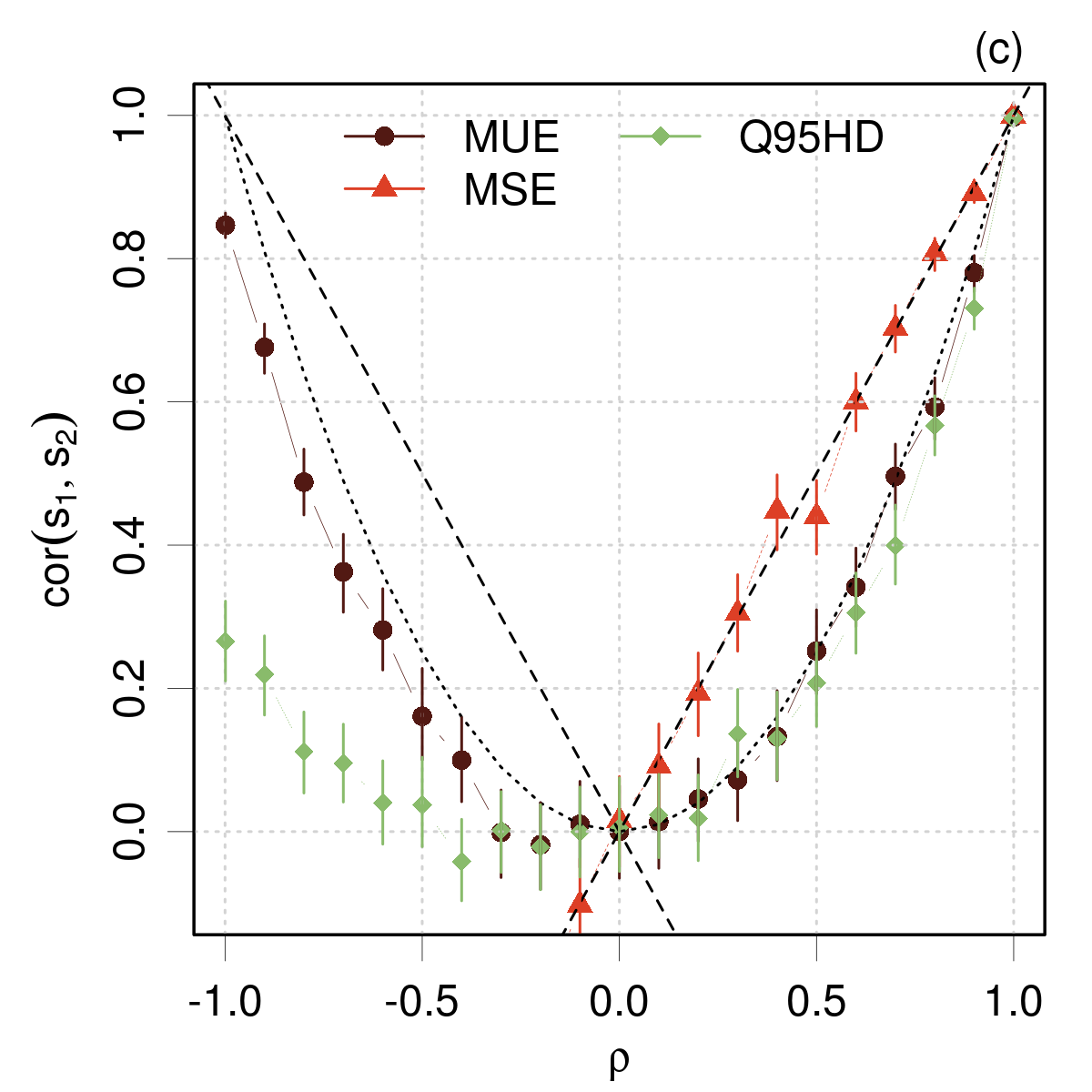}\includegraphics[width=0.32\textwidth]{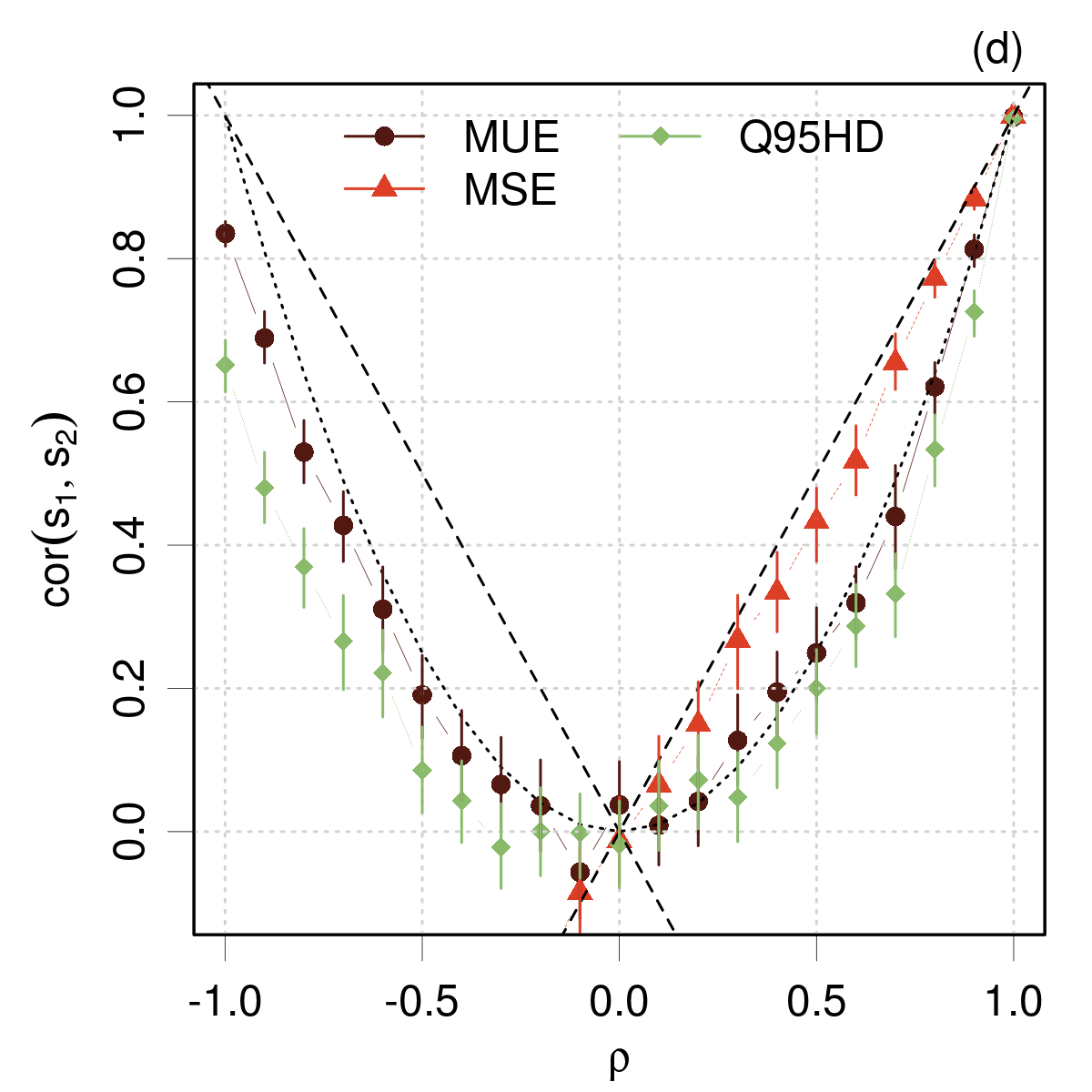}
\par\end{centering}
\noindent \begin{centering}
\includegraphics[width=0.32\textwidth]{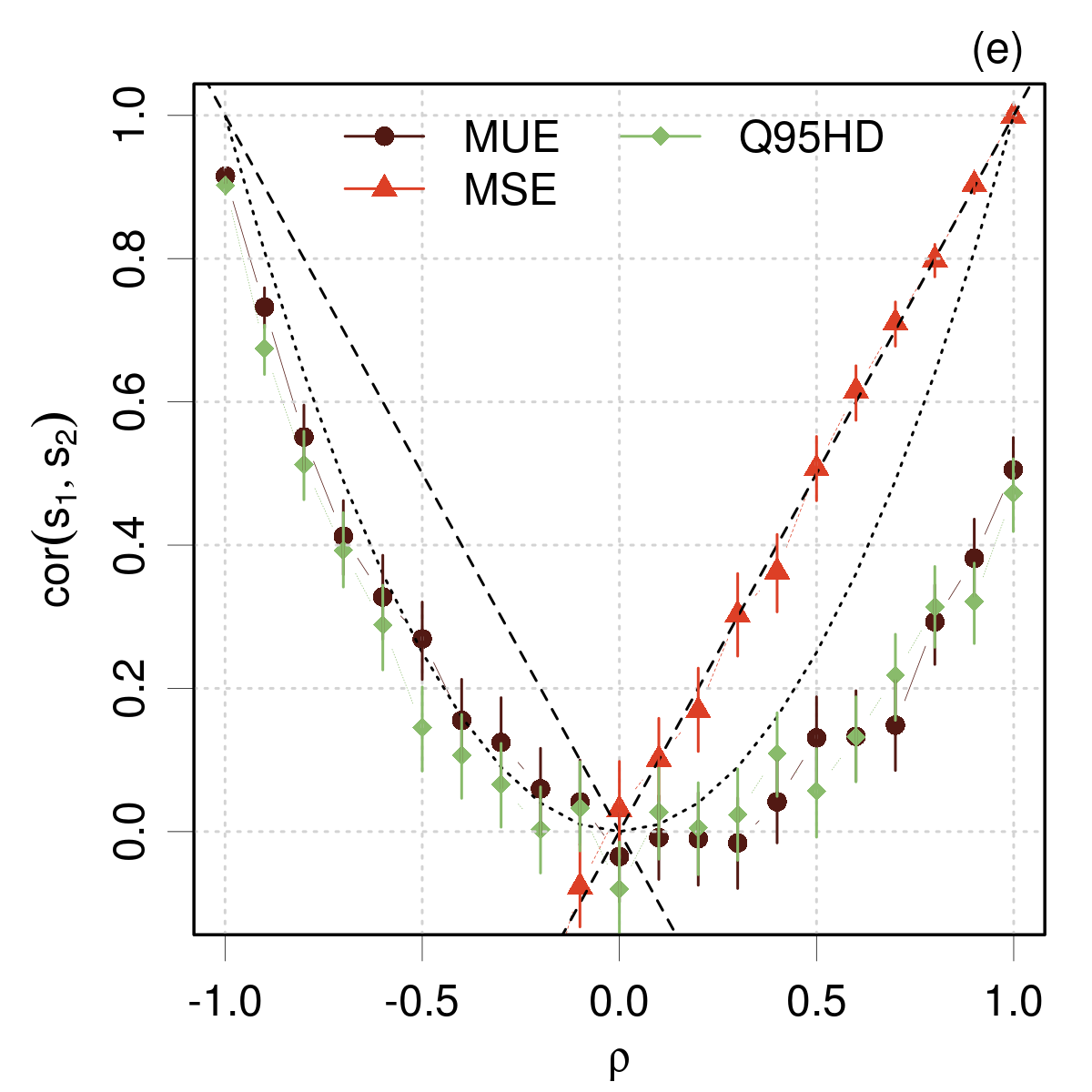}\includegraphics[width=0.32\textwidth]{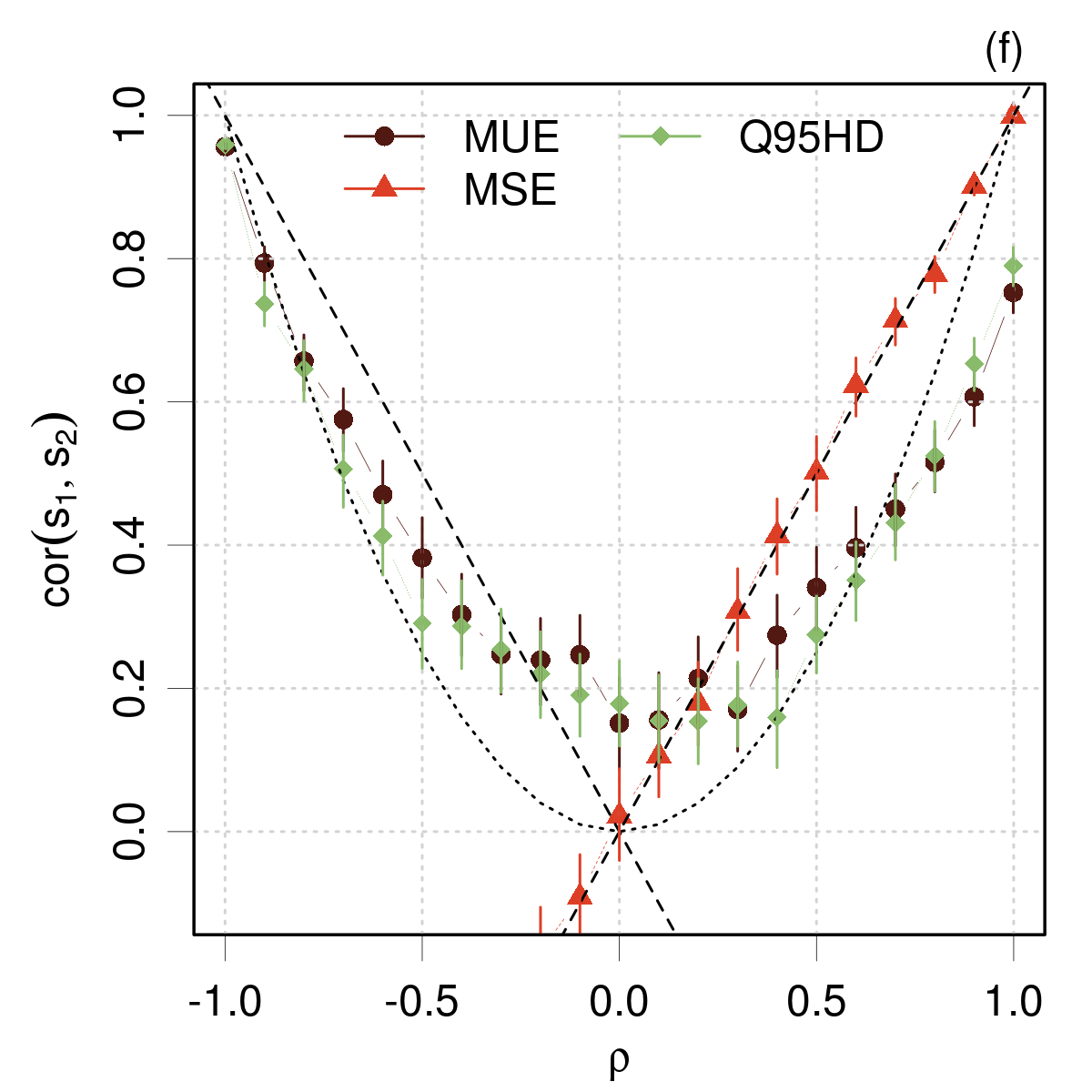}
\par\end{centering}
\noindent \centering{}\caption{\label{fig:corrScore}Correlation coefficients $\mathrm{cor}(s_{1},s_{2})$
of statistics ($S=$\,MUE, MSE, $Q_{95})$ for two samples as a function
of the correlation coefficient $\rho$ of these samples. The error
bars represent 95\,\% intervals for sampling errors. Four cases of
the g-and-h distribution are considered for the error sets: (a) normal
($g=h=0$); (b) heavy-tailed symmetric ($g=0;\thinspace h=0.2$);
(c) light-tailed asymmetric ($g=0.2;\thinspace h=0$); (d) heavy-tailed
asymmetric ($g=h=0.2$). Additional cases with shifted distributions,
$\mu$= (-0.2,0.5) : (e) Normal ; (f) Student's-$t$ ($\nu$= 5).
All distributions have unit variance.}
\end{figure}

The results for four representative cases of the g-and-h distribution
used by Wilcox and Erceg-Hurn \citep{Wilcox2012} (Appendix\,\ref{sec:The-g-and-h-distribution})
of error sets are reported in Fig.\,\ref{fig:corrScore}(a-d). In
this example, both error sets $E_{1}$ and $E_{2}$ have the same
distribution with unit variance, only their correlation varies. 

These simulations confirm the full correlation transfer to the MSE,
independently of the underlying distribution. The correlation coefficients
for the other, non-linear, statistics are mostly positive (within
numerical uncertainty) and systematically smaller than $|\rho|$.
They are symmetrical with respect to $\rho=0$ for symmetrical error
distributions. The values for the MUE are consistently larger than,
or equal to, the values for $Q_{95}$. In all cases, the correlation
coefficient for the MUE is very close to $\rho^{2}$. For negative
values of $\rho$, the correlation coefficient of $Q_{95}$ is sensitive
to the asymmetry or the errors distribution.

The same procedure has been applied to shifted means ($\overline{e}_{1}=-0.2$,
$\overline{e}_{2}=0.5$) for normal and Student's-$t$ distribution
with 5 degrees of freedom (Fig.\,\ref{fig:corrScore}(e,f)). For
the normal distribution the symmetry observed above is broken, as
well as the pure quadratic trend for the MUE. For the Student's-$t$
distribution, the correlations lie above a positive threshold and
one can have $\mathrm{cor}(s_{1},s_{2})>|\rho|$. 

Simulation of correlated error samples enabled us to illustrate properties
of correlation transfer to statistics: identical correlation for the
MSE, and smaller, mostly positive, correlations for the MUE and $Q_{95.}$.
As we covered only a limited set of scenarii, these features cannot
be considered as universal. 

\section{Type I error Probabilities of for the comparison of MUE and $Q_{95}$
pairs\label{sec:Type-I-error}}

A false positive (type I error) is obtained when a true null hypothesis
is rejected by a test \citep{Gregory05a,Klauenberg2019a}. Type I errors
can be kept at a minimum by choosing appropriate data set sizes. Wilcox
and Erceg-Hurn \citep{Wilcox2012} estimated the probability of type
I errors for the comparison of quantiles of correlated data sets with
their method M (Algorithm\,\ref{alg:methodM}) and determined the
sample size $N$ required to reach a probability of type I errors
$\hat{\alpha}$ close to the statistical testing threshold. For their
study, the authors used the g-and-h distribution (Appendix\,\ref{sec:The-g-and-h-distribution})
to generate the data samples, and compared quantiles up two 0.9 for
two levels of correlation, $\rho=0$ and $0.7$. In these conditions,
they concluded that $N\ge30$ was necessary to achieve a correct level
of type I error, considering that it should not exceed $0.075$ for
a test at the 0.05 level \citep{Bradley1978}.

As these test cases did not include our conditions of interest in
terms of correlation (often above $\rho=0.9$) and quantile level
(0.95 for $Q_{95}$), we performed new simulations, using the same
procedure and functions provided in \texttt{R} packages \texttt{WRS}
\citep{R-WRS} and \texttt{WRS2} \citep{R-WRS2}. After assessing
the reproducibility of the original results, we kept the same generative
distribution and scenarii for $g$ and $h$ parameters, and we extended
the exploration for dataset size from $N=20$ to 70, and correlation
coefficient $\rho=0,\thinspace0.5,\thinspace0.9$. 

\begin{figure}[!tb]
\noindent \begin{centering}
\includegraphics[width=0.49\textwidth]{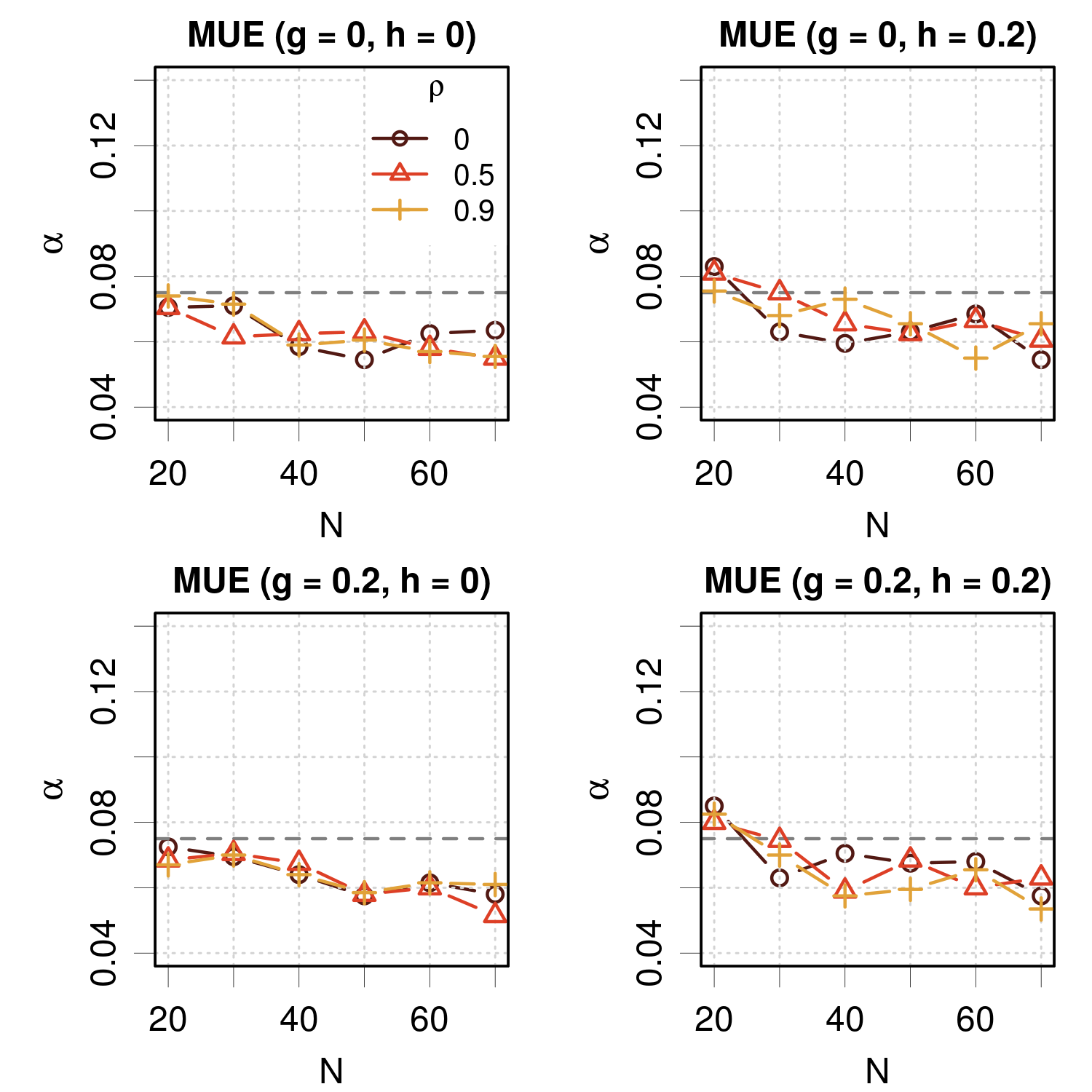}\includegraphics[width=0.49\textwidth]{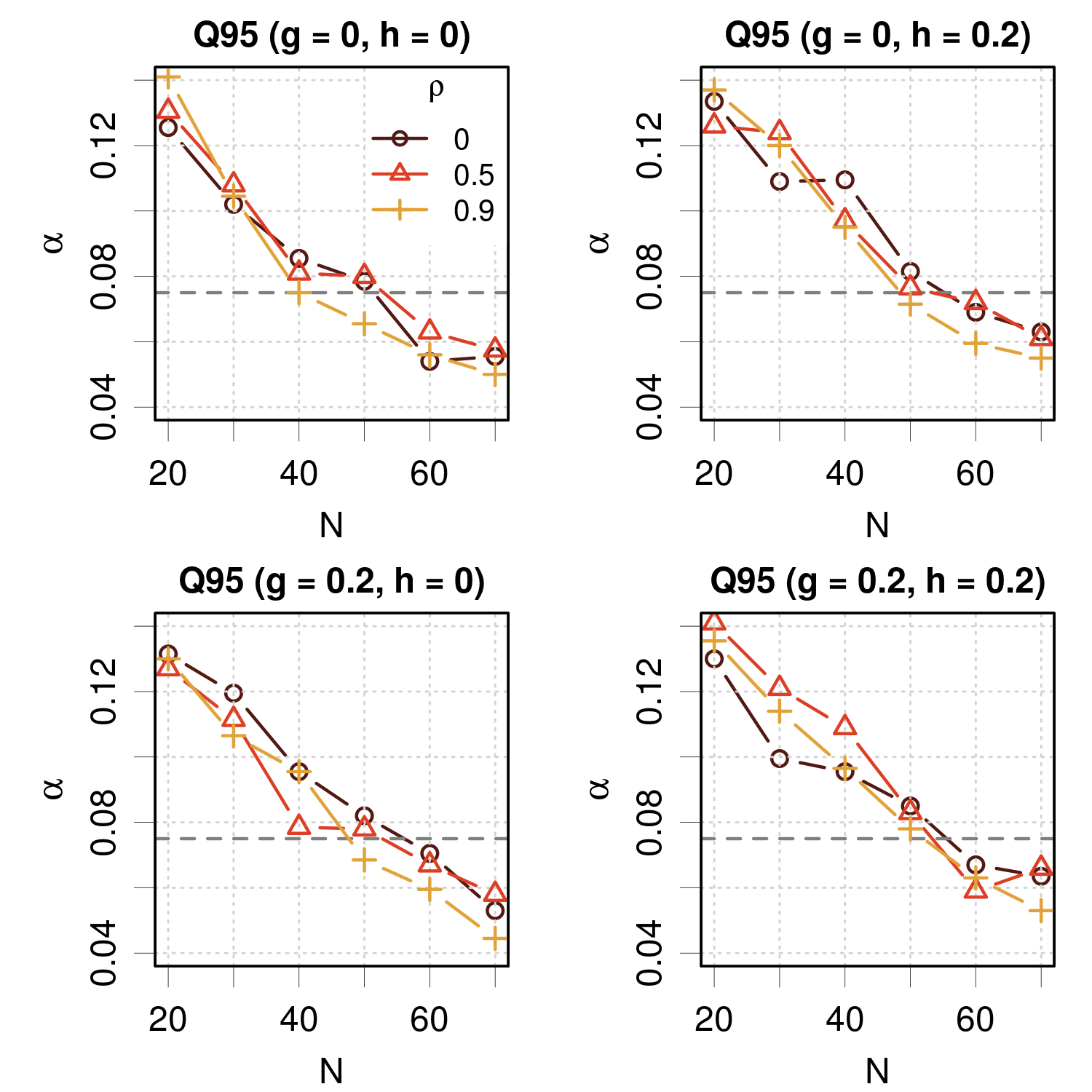}
\par\end{centering}
\noindent \centering{}\caption{\label{fig:power1}Probability of type I errors $\alpha$ for the
MUE (left) and $Q_{95}$ (right), as a function of dataset size $N$.
Each graph corresponds to a type of g-and-h distribution for the data
samples (see text for details). The points and lines correspond to
a value of the datasets correlation coefficient $\rho$. The black
dashed line depicts the upper safety limit (0.075). }
\end{figure}

The procedure is the following: one draws two samples $E_{1}$ and
$E_{2}$ of size $N$ from the same distribution and compute $p_{g}$
for the comparison of the values of a statistic S, $s_{1}$ and $s_{2}$,
respectively. A value of $p_{g}<0.05$ leads to the rejection of the
true null hypothesis $s_{1}=s_{2}$. The process is repeated $M$
times, and the proportion of rejections provides an estimation of
the probability $\alpha$ of type I errors. For compatibility with
the original study, the number of replications is kept to $M=2000$,
and the number of bootstrap samples to $B=1000$. The results for
the comparison of MUE and $Q_{95}$ pairs are reported in Fig.\,\ref{fig:power1}.

For the MUE, the safety region ($\alpha\le0.075$; black dashed line)
\citep{Bradley1978} is reached in all cases for $N\ge30$. Above
$N=40$, all values of $\alpha$ are close to the nominal value (0.05).
There is no remarkable trend with respect to the type of g-and-h distribution,
nor the correlation coefficient. We have estimated previously \citep{Pernot2018}
that the MUE is typically located between the 0.5 and 0.75 quantiles,
for which Wilcox and Erceg-Hurn \citep{Wilcox2012} have concluded
that the minimal dataset size is $N\ge30$, which is confirmed here.

For $Q_{95}$, one sees that for $N=40$, the situation is more favorable
for the normal distribution, but in all cases, the recommended limit
is reached for $N\ge60$. Strong correlation coefficients ($\rho=0.9$)
seem also to be more favorable, and one observes a slight deleterious
effect below $N=50$ for heavy-tailed distributions ($h=0.2$). Nevertheless,
even for $N=30$, $\alpha$ does not exceed notably 12\,\% probability
of type I error.

\paragraph{Remark.}

Establishing the power of the test ($1-\beta$), where $\beta$ is
the probability of type II errors (false negative, or the non-rejection
of a false null hypothesis)\citep{Gregory05a} requires the definition
an alternative hypothesis \citep{Klauenberg2019a}. In the present
case, there is a infinity of ways to realize the $s_{1}\ne s_{2}$
alternative, so the power estimation is practically intractable. 

\section{Numerical study of the Harrell and Davis algorithm\label{sec:Simulated-example}}

This example is intended to outline the advantages of Harrell and
Davis (HD) algorithm for quantiles estimation, notably when associated
with bootstrap sampling, as suggested by Wilcox and Erceg-Hurn \citep{Wilcox2012}. 

One considers the values $s_{1}$ and $s_{2}$ of a statistic $S$
for two datasets $E_{1}$ and $E_{2}$, which are drawn from a bivariate
normal distribution 
\begin{equation}
(E_{1},E_{2})\sim\mathcal{N}\left(\boldsymbol{\mu}=(\mu_{1},\mu_{2}),\boldsymbol{\Sigma}=\left(\begin{array}{cc}
\sigma_{1}^{2} & \rho\sigma_{1}\sigma_{2}\\
\rho\sigma_{1}\sigma_{2} & \sigma_{2}^{2}
\end{array}\right)\right)\label{eq:bivnorm}
\end{equation}
where the error samples have different means $(\mu_{1},\mu_{2})$
and variances $(\sigma_{1}^{2},\sigma_{2}^{2})$, and $\mathrm{cov}(E_{1},E_{2})=\rho\sigma_{1}\sigma_{2}$.
The values of the parameters for the simulations and the corresponding
statistics are given in Table\,\ref{tab:Exact-values}. The reference
values for the MUE and $Q_{95}$ are obtained as described in a previous
article\,\citep{Pernot2018}, based on the properties of the folded
normal distribution.
\begin{table}[!tb]
\begin{centering}
\begin{tabular}{cccccc}
\hline 
Set &  & MSE & RMSD & MUE & $Q_{95}$\tabularnewline
\cline{1-1} \cline{3-6} 
$E_{1}$ &  & 0 & 1.1 & 0.88 & 2.16\tabularnewline
$E_{2}$ &  & 0.1 & 1.0 & 0.80 & 1.97\tabularnewline
\hline 
\end{tabular}
\par\end{centering}
\caption{\label{tab:Exact-values}Reference values for the univariate statistics
of datasets $E_{1}$ and $E_{2}$ described by Eq.\,\ref{eq:bivnorm},
for $\mu_{1}=0$, $\mu_{2}=0.1$, $\sigma_{1}=1.1$ and $\sigma_{2}=1.0$. }
\end{table}

\subsection{Comparison of HD and $\hat{Q}_{7}$ quantiles \label{subsec:Quantiles-estimation-by}}

$Q_{95}$ is estimated by two algorithms: the HD algorithm and the
$\hat{Q}_{7}$ method of Hyndman and Fan \citep{Hyndman1996}, which
is the default algorithms in the \texttt{quantile()} function of \texttt{R}
\citep{CiteR}. $\hat{Q}_{7}$ is one of a family of quantile estimators
based on the linear combination of one or two order statistics \citep{Hyndman1996},
whereas the HD algorithm is based on the linear combination of all
order statistics for a sample \citep{Harrell1982}. The latter is
more efficient for small samples, but more computationally demanding
\citep{Harrell1982}. 

\begin{figure}[!tb]
\noindent \begin{centering}
\includegraphics[width=0.9\columnwidth]{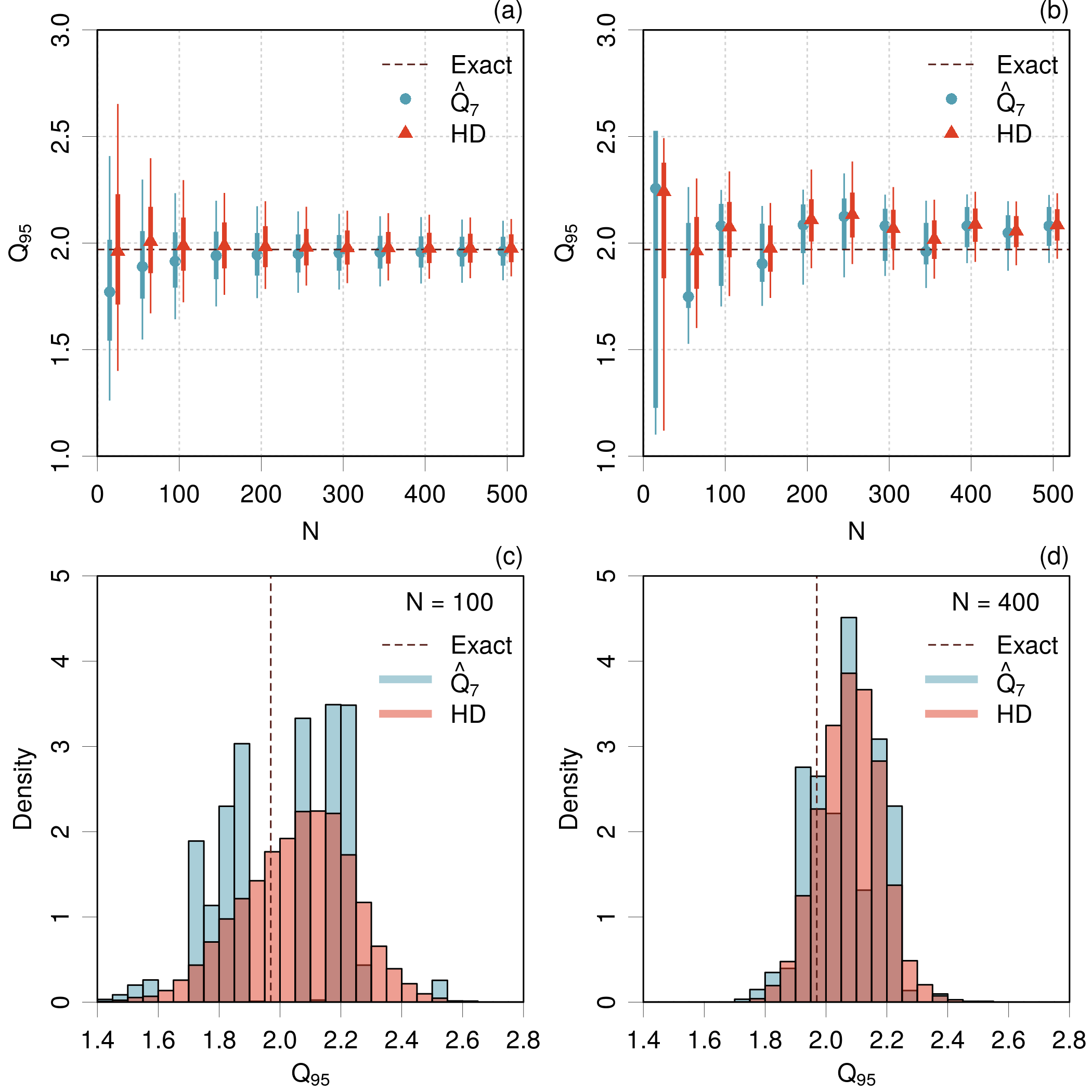}
\par\end{centering}
\noindent \centering{}\caption{\label{fig:simul1}Comparison of $Q_{95}$ estimation algorithms,
$\hat{Q}_{7}$ and HD: (a) Monte Carlo sampling; (b) bootstrap sampling;
(c) bootstrap sample histogram for $N=100$; (d) idem for $N=400$.
The thicker bars in (a,b) represent 25-75\,\% probability intervals
and the finer bars 5-95\,\% probability intervals. The black dashed
line represents the theoretical value for $Q_{95}$ (1.97).}
\end{figure}

In a first test, data sets of increasing sizes, between $N=20$ and
500, are generated by random sampling from the normal distribution
for $E_{2}$, and $Q_{95}$ is estimated for each sample by both algorithms.
This procedure is repeated $10^{4}$ times, and the distributions
of $Q_{95}$ values are summarized by a set of five quantiles (0.05,
0.25, 0.5, 0.75, 0.95). The results are presented in Fig.\,\ref{fig:simul1}(a).
This simulation shows that the HD quantiles converge faster to the
true value (1.97) than the $\hat{Q}_{7}$ ones, with less bias for
small samples ($N<100$). 

In a second test, a unique $E_{2}$ sample of size $N=500$ is generated,
and subsets of increasing size are taken as initial data for a bootstrap
procedure ($10^{4}$ repeats). The bootstrap samples are analyzed
as above and plotted in Fig.\,\ref{fig:simul1}(b). The difference
of convergence between both quantile algorithms is less striking,
but bootstrap for the $\hat{Q}_{7}$ algorithm seems to produce very
asymmetric distributions, where the median is close to one of the
quartiles. If one looks at the histograms of sampled values for $N=100$
(Fig.\,\ref{fig:simul1}(c)), one sees that the HD algorithms produces
a much smoother bootstrap sample histogram, where $\hat{Q}_{7}$ produces
a ragged histograms. The same features are still visible, to a lesser
extent, for $N=400$ (Fig.\,\ref{fig:simul1}(d)). This property
of the HD method explains its good performances for small samples,
when used in conjunction with the bootstrap \citep{Wilcox2012}. 

\subsection{Estimation of $p$-values\label{subsec:Estimation-of--values}}

\begin{figure}[!tb]
\noindent \begin{centering}
\includegraphics[width=0.9\columnwidth]{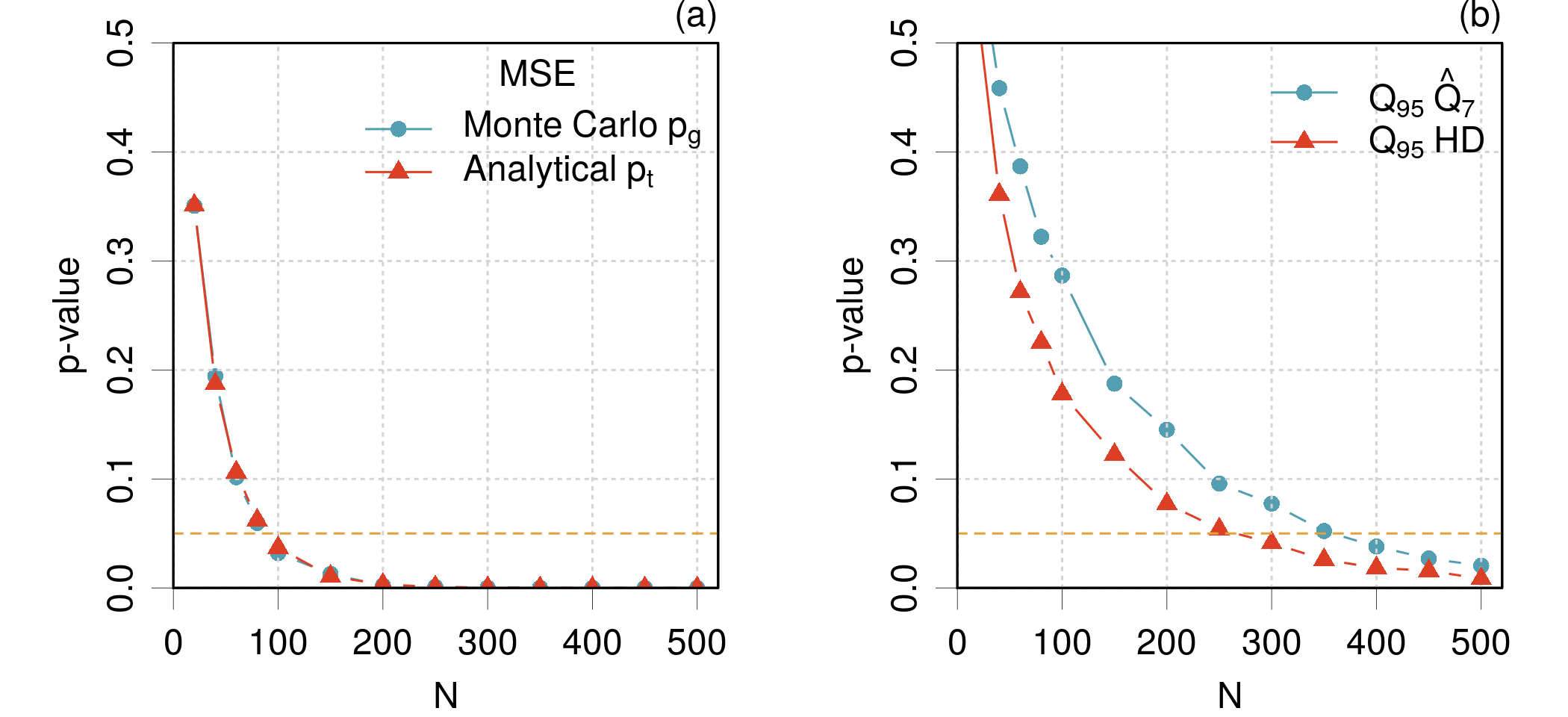}
\par\end{centering}
\noindent \centering{}\caption{\label{fig:scoreBS}Validation of methodological choices for $p$-value
estimation: (a) generalized $p$-value $p_{g}$ for the comparison
of means (MSE) compared to the analytical result $p_{t}$ ; (b) impact
of the quantile estimation algorithm on $p_{g}$ for the comparison
of $Q_{95}$ values. See text for details about the HD and $\hat{Q_{7}}$
algorithm.}
\end{figure}

The estimation of $p$-values is obtained by Monte Carlo sampling
of $E_{1}$ and $E_{2}$ sets of size $N$ varying between 20 and
500 ($\rho=0.9$). One first checks that the generalized $p$-value
$p_{g}$ (Algorithm\,\ref{alg:methodM}) is identical to the analytical
value of $p_{t}$ for the comparison of mean values (Fig.\,\ref{fig:scoreBS}(a)). 

Then, the interest of the Harrell-Davis algorithm for the estimation
of $p_{g}$ values for the comparison of quantiles is shown in Fig.\,\ref{fig:scoreBS}(b):
reaching the 0.05 threshold requires about 250 points for the HD method,
whereas the $\hat{Q}_{7}$ reference quantile algorithm requires about
380 points. Besides, the HD curve is smoother than the reference one,
due to the smoothness properties of the HD estimator shown above.

\section{The g-and-h distribution\label{sec:The-g-and-h-distribution}}

The g-and-h distribution \citep{Hoaglin1985} is typically used to
study the impact of distribution shapes on statistics. If $z$ has
a standard normal distribution, its transform
\begin{equation}
X=\begin{cases}
\frac{1}{g}(e^{gz}-1)e^{\frac{h}{2}z^{2}}, & \mathrm{if}\thinspace g>0\\
ze^{\frac{h}{2}z^{2}} & \mathrm{if}\thinspace g=0
\end{cases}
\end{equation}
has a g-and-h distribution. Its shape is defined by parameters $g$
and $h$, and contains the normal distribution as a special case ($g=h=0$).
Besides the normal, three typical cases are proposed by Wilcox and
Erceg-Hurn \citep{Wilcox2012}: heavy-tailed symmetric ($g=0;\thinspace h=0.2$),
light-tailed asymmetric ($g=0.2;\thinspace h=0$), and heavy-tailed
asymmetric ($g=h=0.2$). 

\bibliographystyle{unsrturlPP}
\phantomsection\addcontentsline{toc}{section}{\refname}\bibliography{packages,NN}

\end{document}